%% file: emanu.tex
\documentclass[11pt,a4paper]{article}
\pdfoutput=1
\usepackage{jcappub}
\usepackage{amsmath}
\usepackage{color}
\usepackage{natbib}
\usepackage{ctable}
\usepackage{bm}
\usepackage{xspace}
\usepackage{csvsimple} 

\let\oldbibliography\thebibliography 
\renewcommand{\thebibliography}[1]{%
  \oldbibliography{#1}%
  \setlength{\itemsep}{0pt}%
  \setlength{\parsep}{0pt}%
  \setlength{\parskip}{0pt}%
  \setlength{\bibsep}{0ex}
  \raggedright
}
\newcommand{\Om}{\Omega_{\rm m}} 
\newcommand{\Ob}{\Omega_{\rm b}} 
\newcommand{\OL}{\Omega_\Lambda}
\newcommand{\smnu}{M_\nu}
\newcommand{\sig}{\sigma_8} 
\newcommand{\mmin}{M_{\rm min}}
\newcommand{\BOk}{\widehat{B}_0} 
\newcommand{\hmpc}{\,h/\mathrm{Mpc}}
\newcommand{\bfi}[1]{\textbf{\textit{#1}}}
\newcommand{\parti}[1]{\frac{\partial #1}{\partial \theta_i}}
\newcommand{\partj}[1]{\frac{\partial #1}{\partial \theta_j}}
\newcommand{\mpc}{h/{\rm Mpc}}

\newcommand{\bitem}{\begin{itemize}}
\newcommand{\eitem}{\end{itemize}}
\newcommand{\beq}{\begin{equation}}
\newcommand{\eeq}{\end{equation}}

\definecolor{dred}{rgb}{0.59,0.,0.09}

\title{Constraining $\smnu$ with the Bispectrum I: Breaking Parameter Degeneracies}

\author[a,b]{ChangHoon Hahn \note{}}
\emailAdd{hahn.changhoon@gmail.com}
\author[c,d]{Francisco Villaescusa-Navarro} 
\author[a,b]{Emanuele Castorina} 
\author[e]{Roman Scoccimarro} 

\affiliation[a]{Lawrence Berkeley National Laboratory, 1 Cyclotron Rd, Berkeley CA 94720, USA}
\affiliation[b]{Berkeley Center for Cosmological Physics, University of California, Berkeley CA 94720, USA}
\affiliation[c]{Center for Computational Astrophysics, Flatiron Institute, 162 5th Avenue, New York NY 10010, USA} 
\affiliation[d]{Department of Astrophysical Sciences, Princeton University, Peyton Hall, Princeton NJ 08544, USA} 
\affiliation[e]{Center for Cosmology and Particle Physics, Department of Physics, New York University, New York, NY 10003 USA}

\abstract{
    Massive neutrinos suppress the growth of structure below their free-streaming scale and leave an 
    imprint on large-scale structure. Measuring this imprint allows us to constrain the sum of neutrino 
    masses, $\smnu$, a key parameter in particle physics beyond the Standard Model. However, degeneracies 
    among cosmological parameters, especially between $\smnu$ and $\sig$, limit the constraining power 
    of standard two-point clustering statistics. In this work, we investigate whether we can break these 
    degeneracies and constrain $\smnu$ with the next higher-order correlation function --- the bispectrum. 
    We first examine the redshift-space halo bispectrum of $800$ $N$-body simulations from the HADES suite 
    and demonstrate that the bispectrum helps break the $\smnu$--$\sig$ degeneracy. Then using 22,000
    $N$-body simulations of the Quijote suite, we quantify for the first time the full information content 
    of the redshift-space halo bispectrum down to nonlinear scales using a Fisher matrix forecast of 
    $\{\Om$, $\Ob$, $h$, $n_s$, $\sig$, $\smnu\}$. For $k_{\rm max}{=}0.5~\mpc$, the bispectrum provides 
    $\Om$, $\Ob$, $h$, $n_s$, and $\sig$ constraints 1.9, 2.6, 3.1, 3.6, and 2.6 times tighter than the 
    power spectrum. For $\smnu$, the bispectrum improves the 1$\sigma$ constraint from 
    0.2968 to 0.0572 eV --- over 5 times tighter than the power spectrum. Even with priors from {\em Planck}, 
    the bispectrum improves $\smnu$ constraints by a factor of 1.8. Although we reserve marginalizing 
    over a more complete set of bias parameters to the next paper of the series, these constraints 
    are derived for a $(1~h^{-1}{\rm Gpc})^3$ box, a substantially smaller volume than upcoming surveys. 
    Thus, our results demonstrate that the bispectrum offers significant improvements over the power 
    spectrum, especially for constraining $\smnu$.  
}

\begin{document}
\maketitle
\flushbottom

\input{intro.tex}

\input{sims.tex}

\input{bk.tex}

\section{Results} \label{sec:results} 
\input{hades.tex} 
\input{quijote.tex}

\input{summary.tex}

\section*{Acknowledgements}
It's a pleasure to thank 
    Michael R. Blanton, 
    William Coulton, 
    Enea Di Dio, 
    Daniel Eisenstein, 
    Simone Ferraro, 
    Simon Foreman, 
    Patrick McDonald,
    Shirley Ho, 
    Emmanuel Schaan, 
    Uro{\u s}~Seljak,
    Zachary Slepian, 
    David N.~Spergel, 
    Licia Verde, 
    and Benjamin D.~Wandelt
    for valuable discussions and comments. 
This material is based upon work supported by the U.S. Department 
of Energy, Office of Science, Office of High Energy Physics, under 
contract No. DE-AC02-05CH11231.
This project used resources of the National Energy Research 
Scientific Computing Center, a DOE Office of Science User 
Facility supported by the Office of Science of the U.S. 
Department of Energy under Contract No. DE-AC02-05CH11231.

\appendix
\input{rsd.tex} 
\input{converge.tex}

\bibliographystyle{yahapj}
\bibliography{emanu} 
\end{document}

%% file: intro.tex
\section{Introduction}
The lower bound on the sum of neutrino masses ($\smnu \gtrsim 0.06$ eV), 
discovered by neutrino oscillation experiments, provides conclusive evidence of 
physics beyond the Standard Model of particle physics~\citep{forero2014, gonzalez-garcia2016}. 
A more precise measurement of $\smnu$ has the potential to distinguish 
between the `normal' and `inverted' neutrino mass hierarchy scenarios 
and further reveal the physics of neutrinos. Neutrino oscillation 
experiments, however, are insensitive to the absolute neutrino mass scales. 
Other laboratory experiments sensitive to $\smnu$ (\emph{e.g.} double beta 
decay and tritium beta decay experiments) have the potential to place upper 
bounds of $\smnu < 0.2$ eV in upcoming experiments~\citep{bonn2011, drexlin2013}. 
However, these upper bound alone are not sufficient to distinguish between the mass
hierarchies. Neutrinos, through the cosmic neutrino background, affect the 
expansion history and the growth of cosmic structure. Measuring these effects 
with cosmological observables provides complementary and potentially more 
precise measurements of $\smnu$. 

Neutrinos, in the early Universe, are relativistic and contribute to the 
energy density of radiation. Later as they become non-relativistic, 
they contribute to the energy density of matter. This transition affects 
the expansion history of the Universe and leaves imprints observable in 
the cosmic microwave background (CMB) anisotropy spectrum~\citep{lesgourgues2012, lesgourgues2014}. 
Massive neutrinos also impact the growth of structure. On large scales, 
neutrino perturbations are indistinguishable from perturbations of cold 
dark matter (CDM). However, on scales smaller than their free-streaming 
scale, neutrinos do not contribute to the clustering and thereby reduce 
the amplitude of the total matter power spectrum. In addition, they also reduce the growth 
rate of CDM perturbations at late times. This combined suppression of 
the small-scale matter power spectrum leaves measurable imprints 
on the CMB as well as large-scale structure. For more on details the effect
of neutrinos in cosmological observables, we refer readers to 
\cite{lesgourgues2012,lesgourgues2014} and~\cite{gerbino2018}. 

The tightest cosmological constraints on $\smnu$ currently come from 
combining CMB data with other cosmological probes. Temperature and large 
angle polarization data from the {\em Planck} satellite places an upper 
bound of $\smnu < 0.54$ eV with 95\% confidence level~\citep{planckcollaboration2018}. 
Adding the Baryon Acoustic Oscillation (BAO) to the {\em Planck} 
likelihood breaks geometrical degeneracies (among $\smnu$, $h$, $\Om$) 
and significantly tightens the upper bound to $\smnu < 0.16$ eV. CMB 
lensing further tightens the bound to $\smnu < 0.13$ eV, though 
not as significantly. Future improvements will likely continue to come from combining CMB data 
on large scales with clustering/lensing data on small scales and low 
redshifts, where the suppression of power by neutrinos is strongest~\citep{brinckmann2019}. 
CMB experiments, however, measure the combined quantity $A_s e^{-2\tau}$, 
where $\tau$ is the optical depth of reionization. Hence, improvements in 
neutrino mass constraints obtained from comparing the power spectrum 
on small and large scales will heavily rely on a better determination of
$\tau$~\citep{allison2015, liu2016, archidiacono2017}. The best constraints on $\tau$ 
currently come from {\em Planck} --- $\tau = 0.054\pm0.007$. However, most
upcoming ground-based CMB experiments ({\em e.g.} CMB-S4) will not observe 
scales larger than $\ell < 30$, and therefore will not directly constrain 
$\tau$~\citep{abazajian2016}. While the upcoming CLASS experiment aims to 
improve $\tau$ constraints~\citep{watts2018}, proposed future space-based 
experiments such as LiteBIRD\footnote{http://litebird.jp/eng/} and 
LiteCOrE\footnote{http://www.core-mission.org/}, which have the greatest 
potential to precisely measure $\tau$, have yet to be confirmed. 
CMB data, however, is not the only way to improve $\smnu$ constraints. The 
imprint of neutrinos on 3D clustering of galaxies can be measured to constrain 
$\smnu$ and with the sheer cosmic volumes mapped, upcoming surveys such 
as DESI\footnote{https://www.desi.lbl.gov/}, PFS\footnote{https://pfs.ipmu.jp/}, 
EUCLID\footnote{http://sci.esa.int/euclid/}, and WFIRST\footnote{https://wfirst.gsfc.nasa.gov/} 
will be able tightly constrain 
$\smnu$~\citep{audren2013, font-ribera2014, petracca2016, sartoris2016, boyle2018}.

A major limitation of using 3D clustering is obtaining accurate theoretical 
predictions beyond linear scales, for bias tracers, and in redshift space. 
Simulations have made huge strides in accurately and efficiently modeling 
nonlinear structure formation with massive neutrinos~\citep[\emph{e.g.}][]{brandbyge2008, 
villaescusa-navarro2013, castorina2015, adamek2017, emberson2017, villaescusa-navarro2018}. 
In conjunction, new simulation based `emulation' models that exploit the 
accuracy of $N$-body simulations while minimizing the computing budget 
have been applied to analyze small-scale galaxy clustering with remarkable 
success~\citep[\emph{e.g.}][]{heitmann2009, kwan2015, euclidcollaboration2018, mcclintock2018, zhai2018, wibking2019}. 
Developments on these fronts have the potential to unlock the information 
content in nonlinear clustering to constrain $\smnu$. 

Various works have examined the impact of neutrino masses on nonlinear clustering 
of matter in 
real-space~\citep[\emph{e.g.}][]{brandbyge2008, saito2008, wong2008, saito2009, viel2010, agarwal2011, bird2012, castorina2015, banerjee2016} 
and in redshift-space~\citep{marulli2011, castorina2015, upadhye2016}. Most recently, 
using a suite of more than 1000 simulations, \cite{villaescusa-navarro2018} 
examined the impact of $\smnu$ on the redshift-space matter and halo power 
spectrum to find that the imprint of $\smnu$ and $\sig$ on the power spectrum are 
degenerate and differ by $< 1\%$ (see also Figure~\ref{fig:plk}). The strong $\smnu$ -- $\sig$ degeneracy
poses a serious limitation on constraining $\smnu$ with the power spectrum. 
However, information in the nonlinear regime cascades from the power spectrum 
to higher-order statistics --- \emph{e.g.} the bispectrum. In fact, the 
bispectrum has a comparable signal-to-noise ratio to the power spectrum
on nonlinear scales~\citep{sefusatti2005, chan2017}. Furthermore, although $\smnu$ 
is not included in their analyses, \cite{sefusatti2006} and \cite{yankelevich2019} 
have shown that including the bispectrum significantly improves constraints on 
cosmological parameters.
Including $\smnu$, \cite{chudaykin2019} find that the bispectrum significantly 
improves constraints for $\smnu$. 
Their forecasts, however, do not include the constraining power on nonlinear scales 
(Section~\ref{sec:forecasts}).
Furthermore, these forecasts do not include the non-Gaussian contributions to the
covariance matrix, which strongly impact the constraining power of the 
bispectrum~\citep{chan2017, sugiyama2019} and require a large number of $N$-body 
simulations to accurately estimate.
No work to date has quantified the total information content and constraining 
power of the full redshift-space bispectrum down to nonlinear scales --- especially for $\smnu$. 

In this work, we examine the effect of massive neutrinos on the redshift-space 
halo bispectrum using ${\sim}23,000$ $N$-body simulations with different neutrino masses. 
We first demonstrate that the bispectrum helps break the $\smnu$--$\sig$ degeneracy 
found in the power spectrum. Then we present the full information content of the 
bispectrum for all triangle configurations down to $k_{\rm max} = 0.5~\mpc$ using 
Fisher forecasts where we estimate the covariance matrix and derivatives with 
the large set of simulations from the Quijote suite~\citep{villaescusa-navarro2019}. 
Afterwards, we explore forecasts with different sets of nuisance and bias parameters 
as well as with and without priors from Planck and discuss some caveats of our 
results. This paper is the first of a series of papers that aim to demonstrate the potential 
of galaxy bispectrum analyses in constraining $\smnu$. In particular, with this 
series, we aim to establish the potential for constraining cosmological parameters 
(especially $\smnu$) with galaxy bispectrum analyses that extend to nonlinear scales 
using simulation based emulation models for upcoming galaxy surveys. 
In this paper, we focus on the halo bispectrum and consider some simple bias models. 
Parameter constraints, however, are derived from the galaxy distribution. In the
subsequent paper, we will include a more realistic galaxy bias model in our forecast 
that describes the distribution of central and satellite galaxies in halos to 
quantify the full information content of the {\em galaxy} bispectrum. 
In the series, we 
will also present methods to tackle challenges that come with analyzing the full galaxy 
bispectrum, such as data compression for reducing the dimensionality of the bispectrum. 

In Section~\ref{sec:hades}, we describe the two $N$-body simulation suites, HADES and Quijote, 
and the halo catalogs constructed from them. We then describe in Section~\ref{sec:bk}, 
how we measure the bispectrum of these simulations. 
Afterwards, we use the redshift-space halo bispectra to demonstrate the distinct imprint of 
$\smnu$ on the bispectrum, which allows it to break the degeneracy between 
$\smnu$ and $\sig$, in Section~\ref{sec:mnusig}. Finally, in Section~\ref{sec:forecasts} 
we present the full information content of the halo bispectrum with a Fisher forecast 
of cosmological parameters and demonstrate how the bispectrum significantly improves 
the constraints on the cosmological parameters: $\Om$, $\Ob$, $h$, $n_s$, $\sig$, and {\em especially} $\smnu$.

%% file: sims.tex
\section{HADES and Quijote Simulation Suites} \label{sec:hades} 
The HADES\footnote{https://franciscovillaescusa.github.io/hades.html} and 
Quijote\footnote{https://github.com/franciscovillaescusa/Quijote-simulations}
suites are sets of, 43000 total, $N$-body simulations run on multiple cosmologies,
including those with massive neutrinos ($\smnu > 0$ eV). In this work, 
we use a subset of the HADES and Quijote simulations. Below, we briefly describe these simulations; 
a summary of the simulations can be found in Table~\ref{tab:sims}. 
The HADES simulations start from Zel'dovich approximated initial conditions 
generated at $z=99$ using the~\cite{zennaro2017a} rescaling method and follow 
the gravitational evolution of $N_{\rm cdm}=512^3$ CDM, plus $N_{\nu}=512^3$ 
neutrino particles for $\smnu > 0$ eV cosmologies, to $z=0$. They are run using 
the {\sc GADGET-III} TreePM+SPH code~\citep{springel2005} in a periodic 
$(1~h^{-1}{\rm Gpc})^3$ box. All of the HADES simulations share the following 
cosmological parameter values, which are in good agreement with Planck 
constraints~\cite{ade2016a}: $\Om{=}0.3175, \Ob{=}0.049, \OL{=}0.6825, n_s{=}0.9624, h{=}0.6711$, 
and $k_{\rm pivot} = 0.05~h{\rm Mpc}^{-1}$. 

The HADES suite includes $N$-body simulations with degenerate massive neutrinos 
of $\smnu = $ 0.06, 0.10, and 0.15 eV. These simulations are run using the 
``particle method'', where neutrinos are described as a collisionless 
and pressureless fluid and therefore modeled as particles, same as 
CDM~\citep{brandbyge2008,viel2010}. HADES also includes simulations with massless 
neutrino and different values of $\sigma_8$ to examine the $\smnu-\sigma_8$ 
degeneracy. The $\sigma_8$ values in these simulations were chosen to match {\em either} 
$\sigma_8^m$ or $\sigma_8^{c}$ ($\sigma_8$ computed with respect to total matter, 
CDM + baryons + $\nu$, or CDM + baryons) of the massive neutrino simulations: 
$\sigma_8 = 0.822, 0.818, 0.807$, and $0.798$. Each model has $100$ independent 
realizations and we focus on the snapshots saved at $z = 0$. Halos closely 
trace the CDM+baryon field rather than the total matter field and neutrinos 
have negligible contribution to halo masses~\citep[\emph{e.g.}][]{ichiki2012, castorina2014, loverde2014, villaescusa-navarro2014}.
Hence, dark matter halos are identified in each realization using the Friends-of-Friends 
algorithm~\cite[FoF;][]{davis1985} with linking length $b=0.2$ on the CDM + baryon
distribution. We limit the catalogs to halos with masses above 
$M_{\rm lim} = 3.2\times 10^{13} h^{-1}M_\odot$. We refer readers
to~\cite{villaescusa-navarro2018} for more details on the HADES simulations. 

In addition to HADES, we use simulations from the Quijote suite, a
set of 42,000 $N$-body simulations that in total contain more than 8 trillion 
($8\times10^{12}$) particles over a volume of $42000 (h^{-1}{\rm Gpc})^3$. 
These simulations were designed to quantify the information content of 
different cosmological observables using Fisher matrix forecasting 
technique~(Section~\ref{sec:forecasts}). They are therefore constructed to accurately 
calculate the covariance matrices of observables and the derivatives of observables with 
respect to cosmological parameters: 
$\Om$, $\Ob$, $h$, $n_s$, $\sig$, and $\smnu$.

To calculate covariance matrices, Quijote includes $N_{\rm cov}{=}15,000$ $N$-body 
simulations at a fiducial cosmology ($\Om{=}0.3175$, $\Ob{=}0.049$, $h{=}0.6711$, 
$n_s{=}0.9624$, $\sig{=}0.834$, and $\smnu{=}0.0$ eV). It also includes sets 
of 500 $N$-body simulations run at different cosmologies where only one parameter 
is varied from the fiducial cosmology at a time for the derivatives. Along $\Om$, 
$\Ob$, $h$, $n_s$, and $\sig$, the fiducial cosmology is adjusted by either a 
small step above or below the fiducial value: 
$\{\Om^{+}, \Om^{-}, \Ob^{+}, \Ob^{-}, h^+, h^-, n_s^{+}, n_s^{-}, \sig^{+}, \sig^-\}$ . 
Along $\smnu$, because $\smnu \ge 0.0$ eV and the derivative of certain observable 
with respect to $\smnu$ is noisy, Quijote includes sets of 500 simulations for 
$\smnu = 0.1$, 0.2, and 0.4 eV. Table~\ref{tab:sims} lists the cosmologies included 
in the Quijote suite. 

The initial conditions for all Quijote simulations were generated at $z=127$ using 
2LPT for simulations with massless neutrinos and the Zel’dovich approximation for 
massive neutrinos. The suite also includes a set of simulations at the fiducial 
$\smnu = 0$ eV cosmology with Zel'dovich approximation initial conditions, which 
we use later in Section~\ref{sec:forecasts}. Like HADES, the initial conditions 
of simulations with massive neutrinos take their scale-dependent growth factors/rates 
into account using the \cite{zennaro2017a} method. From the initial conditions, all 
of the simulations follow the gravitational evolution of $512^3$ dark matter particles, 
and $512^3$ neutrino particles for massive neutrino models, to $z=0$ using {\sc Gadget-III}
TreePM+SPH code (same as HADES). Halos are then identified using the same FoF 
scheme and mass limit as HADES. For the fiducial cosmology, the halo catalogs 
have ${\sim}156,000$ halos ($\bar{n} \sim 1.56 \times 10^{-4}~h^3{\rm Mpc}^{-3}$) 
with $\bar{n} P_0(k=0.1)\sim 3.23$.
We refer readers to~\cite{villaescusa-navarro2019} for further details on the Quijote simulations.

\begin{figure}
\begin{center}
\includegraphics[width=0.9\textwidth]{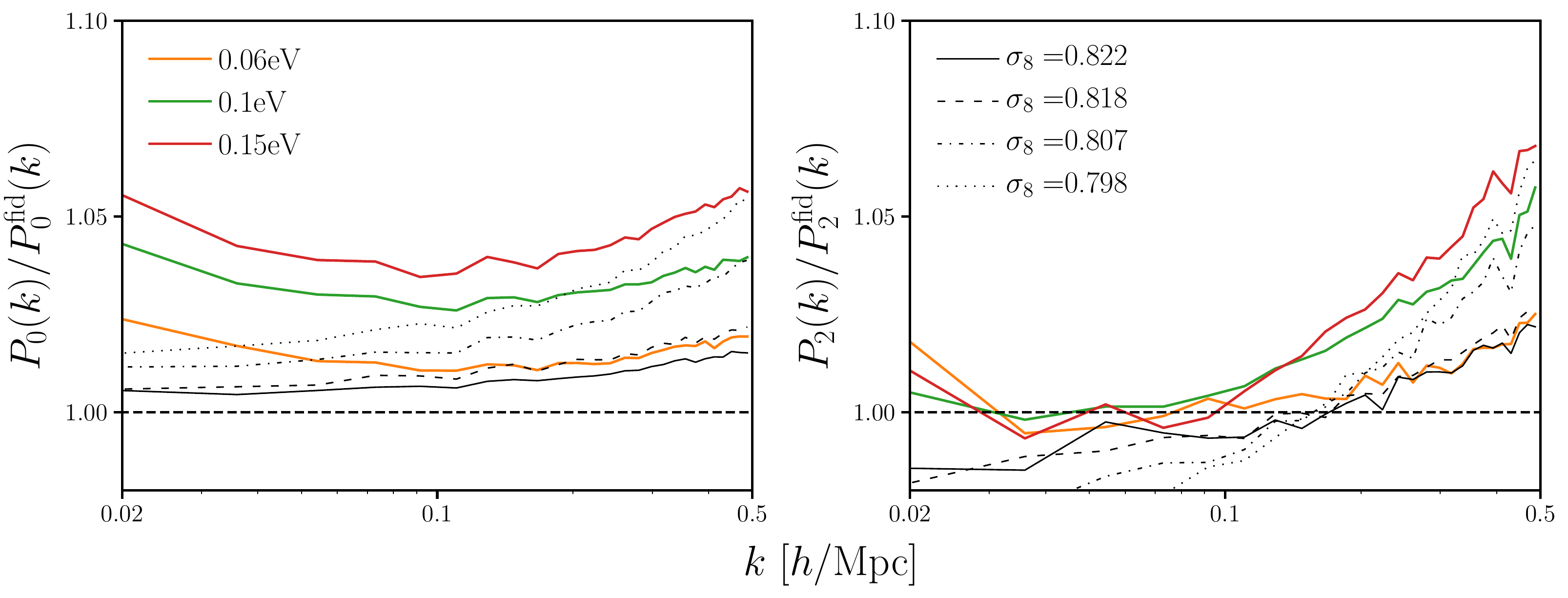}
    \caption{Impact of $\smnu$ and $\sig$ on the redshift-space halo power 
    spectrum monopole, $P_0$, and quadrupole, $P_2$, of the HADES simulation 
    suite, which includes simulations with $\smnu = 0.0$, 0.06, 0.10, 
    and 0.15 eV (orange, green, and red) as well as massless neutrino 
    simulations with $\sig = 0.822$, 0.818, 0.807, and 0.798 
    (black solid, dashed, dot-dashed, and dotted). $P_\ell$ at 
    each of these cosmology is averaged over 100 $N$-body realizations. 
    While neutrinos suppress small-scale matter power spectrum,  
    the halo $P_\ell$ increases with higher $\smnu$ because cosmologies 
    with higher $\smnu$ have lower $\sig$, which translates into larger 
    halo bias for a fixed halo mass limit. 
    $\smnu$ and $\sig$ produce almost identical effects on halo clustering 
    on small scales ($k > 0.1\hmpc$). This degeneracy can be partially 
    broken through the quadrupole; however, {\em $\smnu$ and $\sig$ 
    produce almost the same effect on two-point clustering --- within a 
    few percent}.
    }
\label{fig:plk}
\end{center}
\end{figure}

\begin{table}
\caption{Specifications of the HADES and Quijote simulation suites.} 
\begin{center} 
\resizebox{\textwidth}{!}{\begin{tabular}{cccccccccc} \toprule
    Name  &$\smnu$ & $\Omega_m$ & $\Omega_b$ & $h$ & $n_s$ & $\sigma^m_8$ & $\sigma^c_8$ & ICs & realizations \\
      &({\footnotesize eV}) & & & & & ({\footnotesize $10^{10}h^{-1}M_\odot$}) & ({\footnotesize $10^{10}h^{-1}M_\odot$}) & \\[3pt] \hline\hline
    \multicolumn{9}{c}{HADES suite} \\ \hline
    Fiducial    & 0.0   & 0.3175 & 0.049 & 0.6711 & 0.9624 & 0.833 & 0.833 & Zel'dovich & 100 \\ 
                & 0.06  & 0.3175 & 0.049 & 0.6711 & 0.9624 & 0.819 & 0.822 & Zel'dovich & 100 \\ 
                & 0.10  & 0.3175 & 0.049 & 0.6711 & 0.9624 & 0.809 & 0.815 & Zel'dovich & 100 \\ 
                & 0.15  & 0.3175 & 0.049 & 0.6711 & 0.9624 & 0.798 & 0.806 & Zel'dovich & 100 \\ 
                & 0.0   & 0.3175 & 0.049 & 0.6711 & 0.9624 & 0.822 & 0.822 & Zel'dovich & 100 \\ 
                & 0.0   & 0.3175 & 0.049 & 0.6711 & 0.9624 & 0.818 & 0.818 & Zel'dovich & 100 \\ 
                & 0.0   & 0.3175 & 0.049 & 0.6711 & 0.9624 & 0.807 & 0.807 & Zel'dovich & 100 \\ 
                & 0.0   & 0.3175 & 0.049 & 0.6711 & 0.9624 & 0.798 & 0.798 & Zel'dovich & 100 \\[3pt]
    \hline \hline
    \multicolumn{9}{c}{Qujiote suite} \\ \hline
    Fiducial 	    & 0.0         & 0.3175 & 0.049 & 0.6711 & 0.9624 & 0.834 & 0.834 & 2LPT & 15,000 \\ 
    Fiducial ZA     & 0.0         & 0.3175 & 0.049 & 0.6711 & 0.9624 & 0.834 & 0.834 & Zel'dovich& 500 \\ 
    $\smnu^+$       & \underline{0.1}   & 0.3175 & 0.049 & 0.6711 & 0.9624 & 0.834 & 0.834 & Zel'dovich & 500 \\ 
    $\smnu^{++}$    & \underline{0.2}   & 0.3175 & 0.049 & 0.6711 & 0.9624 & 0.834 & 0.834 & Zel'dovich & 500 \\ 
    $\smnu^{+++}$   & \underline{0.4}   & 0.3175 & 0.049 & 0.6711 & 0.9624 & 0.834 & 0.834 & Zel'dovich & 500 \\ 
    $\Omega_m^+$    & 0.0   & \underline{ 0.3275} & 0.049 & 0.6711 & 0.9624 & 0.834 & 0.834 & 2LPT & 500 \\ 
    $\Omega_m^-$    & 0.0   & \underline{ 0.3075} & 0.049 & 0.6711 & 0.9624 & 0.834 & 0.834 & 2LPT & 500 \\ 
    $\Omega_b^+$    & 0.0   & 0.3175 & \underline{0.051} & 0.6711 & 0.9624 & 0.834 & 0.834 & 2LPT & 500 \\ 
    $\Omega_b^-$    & 0.0   & 0.3175 & \underline{0.047} & 0.6711 & 0.9624 & 0.834 & 0.834 & 2LPT & 500 \\ 
    $h^+$           & 0.0   & 0.3175 & 0.049 & \underline{0.6911} & 0.9624 & 0.834 & 0.834 & 2LPT & 500 \\ 
    $h^-$           & 0.0   & 0.3175 & 0.049 & \underline{0.6511} & 0.9624 & 0.834 & 0.834 & 2LPT & 500 \\ 
    $n_s^+$         & 0.0   & 0.3175 & 0.049 & 0.6711 & \underline{0.9824} & 0.834 & 0.834 & 2LPT & 500 \\ 
    $n_s^-$         & 0.0   & 0.3175 & 0.049 & 0.6711 & \underline{0.9424} & 0.834 & 0.834 & 2LPT & 500 \\ 
    $\sigma_8^+$    & 0.0   & 0.3175 & 0.049 & 0.6711 & 0.9624 & \underline{0.849} & \underline{0.849} & 2LPT & 500 \\ 
    $\sigma_8^-$    & 0.0   & 0.3175 & 0.049 & 0.6711 & 0.9624 & \underline{0.819} & \underline{0.819} & 2LPT & 500 \\[3pt]
    \hline
\end{tabular}} \label{tab:sims}
\end{center}
    {\bf \em Top}: The HADES suite includes sets of 100 $N$-body simulations with degenerate massive neutrinos 
    of $\smnu = $ 0.06, 0.10, and 0.15 eV as well as sets of simulations with $\smnu = 0.0$ eV and 
    $\sigma_8 = $ 0.822, 0.818, 0.807, and 0.798 to examine the $\smnu-\sigma_8$ degeneracy. 
    {\bf \em Bottom}: The Quijote suite includes 15,000 $N$-body simulations at the fiducial 
    cosmology to accurately estimate the covariance matrices. It also includes sets of 500 
    simulations at 13 different cosmologies, where only one parameter is varied from the fiducial 
    value (underlined), to estimate derivatives of observables along the cosmological parameters.
\end{table}

%% file: bk.tex
\begin{figure}
\begin{center}
    \includegraphics[width=\textwidth]{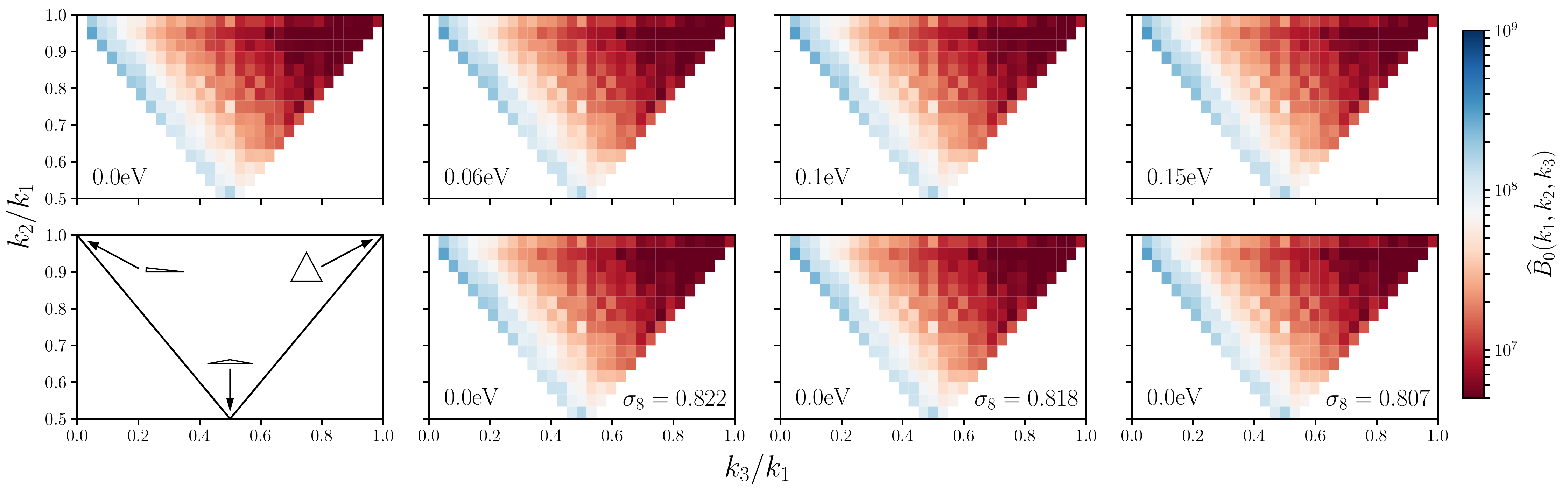} 
    \caption{The redshift-space halo bispectrum, $\BOk(k_1, k_2, k_3)$, as a 
    function of triangle configuration shape for $\smnu = 0.0, 0.06, 0.10$, 
    and $0.15\,\mathrm{eV}$ (upper panels) and $\smnu=0.0$ eV, $\sig = 0.822, 0.818$, and 
    $0.807$ (lower panels). The $\BOk$ for each cosmology (each panel) is 
    averaged over 100 $N$-body realizations. The HADES simulations of the top 
    and bottom panels in the three right-most columns, have matching $\sig$ values 
    (Section~\ref{sec:hades}). We describe the triangle configuration shape 
    by the ratio of the triangle sides: $k_3/k_1$ and $k_2/k_1$. As we describe 
    schematically in the lower leftmost panel, in each panel, the upper left bin 
    contains squeezed triangles ($k_1 = k_2 \gg k_3$), the upper right bin 
    contains equilateral triangles ($k_1 = k_2 = k_3$), and the bottom center bin 
    contains folded triangles ($k_1 = 2 k_2 = 2 k_3$). We include all 1898 triangle 
    configurations with $k_1, k_2, k_3 \leq k_{\rm max} = 0.5~\mpc$ and use the
    $\BOk$ estimator in Section~\ref{sec:bk}.}
\label{fig:bk_shape}
\end{center}
\end{figure}

\begin{figure}
\begin{center}
\includegraphics[width=0.9\textwidth]{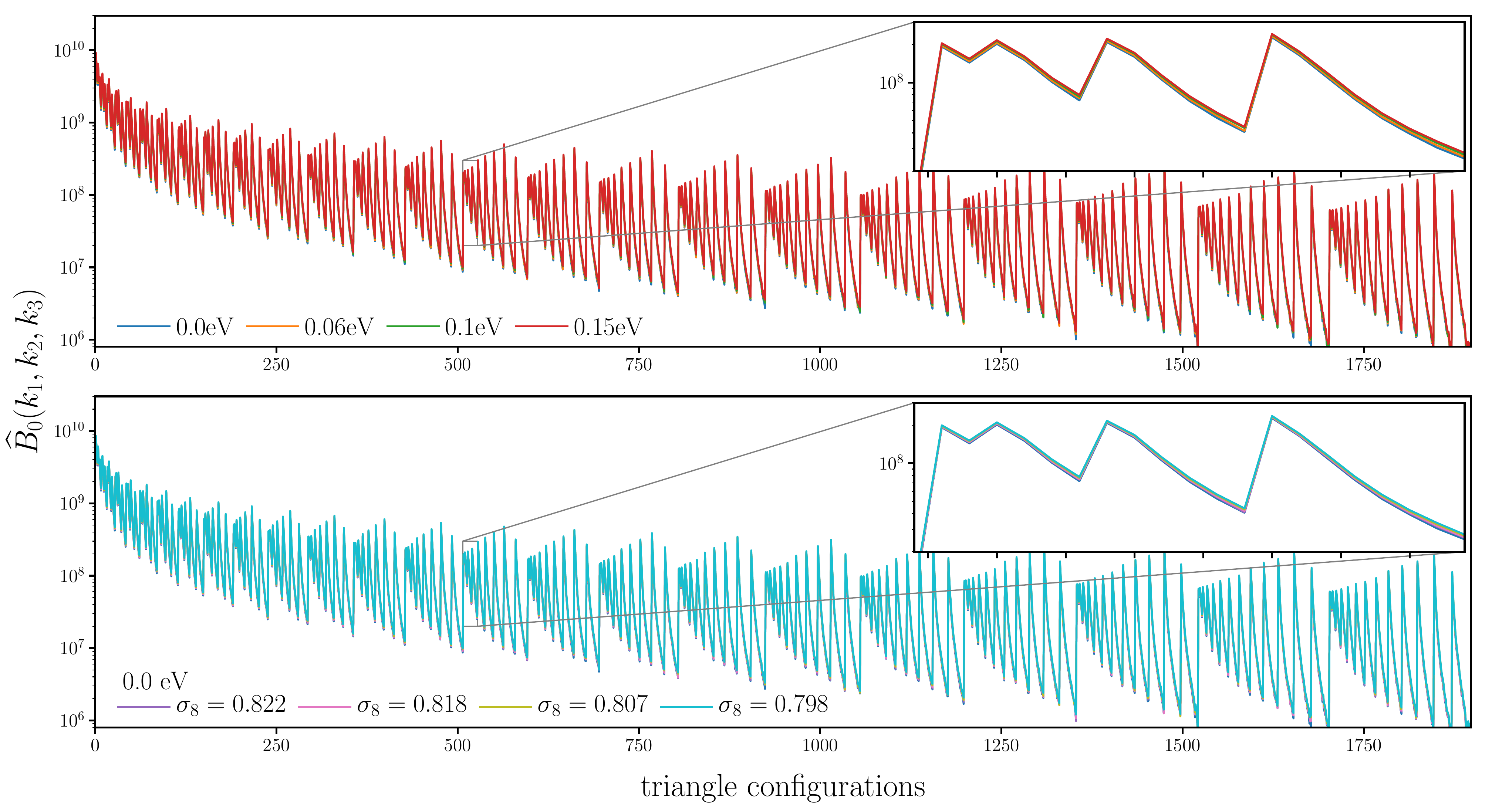}
    \caption{The redshift-space halo bispectrum, $\BOk(k_1, k_2, k_3)$, as a
    function of triangle configurations for $\smnu = 0.0, 0.06, 0.10$, 
    and $0.15\,\mathrm{eV}$ (top panel) and $\smnu = 0.0$ eV, $\sig = 0.822, 0.818, 0.807$, 
    and $0.798$ (lower panel). Each $\BOk$ is averaged over 100 $N$-body 
    realizations. We include all possible triangle configurations 
    with $k_1, k_2, k_3 \leq k_{\rm max} = 0.5~\mpc$ where we order
    the configurations by looping through $k_3$ in the inner most loop and 
    $k_1$ in the outer most loop satisfying $k_1 \geq k_2 \geq k_3$ 
    (see Appendix~\ref{sec:bk_details} for more details). In the 
    insets of the panels we zoom into triangle configurations with 
    $k_1 = 0.320$, $0.170 \leq k_2 \leq 0.302$, and 
    $0.094 \leq k_3 \leq 0.302~h/{\rm Mpc}$.}
\label{fig:bk_amp}
\end{center}
\end{figure}

\section{Bispectrum} \label{sec:bk} 
We are interested in breaking parameter degeneracies that limit the constraining 
power on $\smnu$ of two-point clustering analyses using three-point clustering 
statistics --- \emph{i.e.} the bispectrum. In this section, we describe the 
bispectrum estimator used throughout the paper. We focus on the bispectrum monopole 
($\ell = 0$) and use an estimator that exploits the speed of Fast Fourier 
Transforms (FFTs). Our estimator is similar to the estimators described in 
\cite{sefusatti2005a}, \cite{scoccimarro2015}, and \cite{sefusatti2016}; we also 
follow their formalism in our description below. Although \cite{sefusatti2016} 
and \cite{scoccimarro2015} respectively describe estimators in real- and 
redshift-space, since we focus on the bispectrum monopole, we note that there 
is no difference. 

To measure the bispectrum of our halo catalogs, we begin by interpolating the halo
positions to a grid, $\delta({\bfi x})$, and Fourier transforming the grid to get 
$\delta({\bfi k})$. We use a fourth-order interpolation to get interlaced grids, 
which has advantageous anti-aliasing properties that allow unbiased measurements 
up to the Nyquist frequency~\citep{hockney1981,sefusatti2016}. Then using 
$\delta({\bfi k})$ we measure the bispectrum monopole as 
\beq \label{eq:bk} 
\widehat{B}_{\ell = 0}(k_1,k_2, k_3) = \frac{1}{V_B} \int\limits_{k_1}{\rm d}^3q_1 \int\limits_{k_2}{\rm d}^3q_2 \int\limits_{k_3}{\rm d}^3q_3~\delta_{\rm D}({\bfi q_{123}})~\delta({\bfi q_1})~\delta({\bfi q_2})~\delta({\bfi q_3}) - B^{\rm SN}_{\ell = 0}.
\eeq
$\delta_{\rm D}$ is the Dirac delta function and hence $\delta_{\rm D}({\bfi q_{123}}) = \delta_{\rm D}({\bfi q_1} + {\bfi q_2} + {\bfi q_3})$ 
ensures that the ${\bfi q}_i$ triplet actually forms a closed triangle. Each of the integrals 
above represent an integral over a spherical shell in $k$-space with radius $\delta k$ 
centered at ${\bfi k}_i$: 
\beq
\int_{k_i}{\rm d}^3q \equiv \int\limits_{k_i-\delta k/2}^{k_i+\delta k/2}{\rm d}q~q^2\int {\rm d}\Omega.
\eeq
$V_B$ is a normalization factor proportional to the number of triplets ${\bfi q_1}$, 
${\bfi q_2}$, and ${\bfi q_3}$ that can be found in the triangle bin defined by 
$k_1$, $k_2$, and $k_3$ with width $\delta k$: 
\beq
V_B = \int\limits_{k_1}{\rm d}^3q_1 \int\limits_{k_2}{\rm d}^3q_2 \int\limits_{k_3}{\rm d}^3q_3~\delta_{\rm D}({\bfi q_{123}}). 
\eeq
Lastly, $B^{\rm SN}_{\ell=0}$ is the correction for the Poisson shot noise, which 
contributes due to the self-correlation of individual objects: 
\beq \label{eq:bk_sn} 
B^{\rm SN}_{\ell=0}(k_1, k_2, k_3) = \frac{1}{\bar{n}} \big(P_0(k_1) + P_0(k_2) + P_0(k_3) \big) + \frac{1}{\bar{n}^2}
\eeq
where $\bar{n}$ is the number density of objects (halos) and $P_0$ is the power 
spectrum monopole. 

In order to evaluate the integrals in Eq.~\ref{eq:bk}, we take advantage of the plane-wave 
representation of the Dirac delta function and rewrite the equation as
\begin{align} \label{eq:bk2} 
    \widehat{B}_{\ell = 0}(k_1,k_2, k_3) &= \frac{1}{V_B} \int\frac{{\rm d}^3x}{(2\pi)^3} \int\limits_{k_1}{\rm d}^3q_1 \int\limits_{k_2}{\rm d}^3q_2 \int\limits_{k_3}{\rm d}^3q_3~\delta({\bfi q_1})~\delta({\bfi q_2})~\delta({\bfi q_3})~e^{i{\bfi q_{123}}\cdot{\bfi x}} - B^{\rm SN}_{\ell = 0}\\ 
    &= \frac{1}{V_B} \int\frac{{\rm d}^3x}{(2\pi)^3}\prod\limits_{i=1}^{3} I_{k_i}({\bfi x}) - B^{\rm SN}_{\ell = 0} 
\end{align}
where 
\beq
I_{k_i}({\bfi x}) = \int\limits_k {\rm d}^3q~\delta({\bfi q})~e^{i {\bfi q}\cdot{\bfi x}}. 
\eeq
At this point, we measure $\widehat{B}_{\ell = 0}(k_1, k_2, k_3)$ by calculating the 
$I_{k_i}$s with inverse FFTs and summing over in real space. For $\widehat{B}_{\ell = 0}$
measurements throughout the paper, we use $\delta({\bfi x})$ grids with $N_{\rm grid} = 360$ 
and triangle configurations defined by $k_1, k_2, k_3$ bins of width 
$\Delta k = 3 k_f = 0.01885~h/{\rm Mpc}$, three times the fundamental mode
$k_f = 2\pi/(1000~h^{-1}{\rm Mpc})$ given the box size\footnote{The code that we use to 
evaluate $\widehat{B}_{\ell = 0}$ is publicly available at \url{https://github.com/changhoonhahn/pySpectrum}}. 

We present the redshift-space halo bispectrum of the HADES simulations measured using 
the estimator above in two ways: one that emphasizes the triangle shape dependence 
(Figure~\ref{fig:bk_shape}) and the other that emphasizes the amplitude 
(Figure~\ref{fig:bk_amp}). In Figure~\ref{fig:bk_shape}, we plot $\BOk(k_1, k_2, k_3)$ 
as a function of $k_2/k_1$ and $k_3/k_1$, which describe the triangle configuration shapes. 
In each panel, the colormap of the ($k_2/k_1$, $k_3/k_1$) bins represent the weighted 
average $\BOk$ amplitude of all triangle configurations in the bins. The upper left 
bins contain squeezed triangles ($k_1 = k_2 \gg k_3$), the upper right bins contain 
equilateral triangles ($k_1 = k_2 = k_3$), and the bottom center bins contain folded 
triangles ($k_1 = 2 k_2 = 2 k_3$) as schematically highlighted in the lower leftmost panel. 
We include all possible 1898 triangle configurations with $k_1, k_2, k_3 < k_{\rm max} = 0.5~\mpc$. 
$\BOk$ in the upper panels are HADES models with 
$(\smnu,~\sig) = (0.0~{\rm eV},~0.833)$ (fiducial), $(0.06~{\rm eV},~0.822), (0.10~{\rm eV},~0.815)$, 
and $(0.15~{\rm eV},~0.806)$. $\BOk$ in the lower panels are  HADES models with $\smnu = 0.0$ eV and 
$\sig = 0.822, 0.818$, and $0.807$. The top and bottom panels of the three right-most 
columns have matching $\sig$ values (Section~\ref{sec:hades}).

Next, in Figure~\ref{fig:bk_amp} we plot $\BOk(k_1, k_2, k_3)$ for all possible triangle 
configurations with $k_1, k_2, k_3 < k_{\rm max} = 0.5~\mpc$ where we order the 
configurations by looping through $k_3$ in the inner most loop and $k_1$ in the outer most 
loop with $k_1 \geq k_2 \geq k_3$. In the top panel, we present $\BOk$ of HADES models 
with $\smnu = 0.0, 0.06, 0.10$, and $0.15\,\mathrm{eV}$; in the lower panel, we present 
$\BOk$ of HADES models with $\smnu = 0.0$ eV and $\sig = 0.822, 0.818$, and $0.807$. We 
zoom into triangle configurations with $k_1 = 0.320$, $0.170 \leq k_2 \leq 0.302$, and 
$0.094 \leq k_3 \leq 0.302~h/{\rm Mpc}$ in the insets of the panels. For further details on the
redshift-space bispectrum, we refer to Appendix~\ref{sec:bk_details}.

%% file: hades.tex
\begin{figure}
\begin{center}
\includegraphics[width=0.8\textwidth]{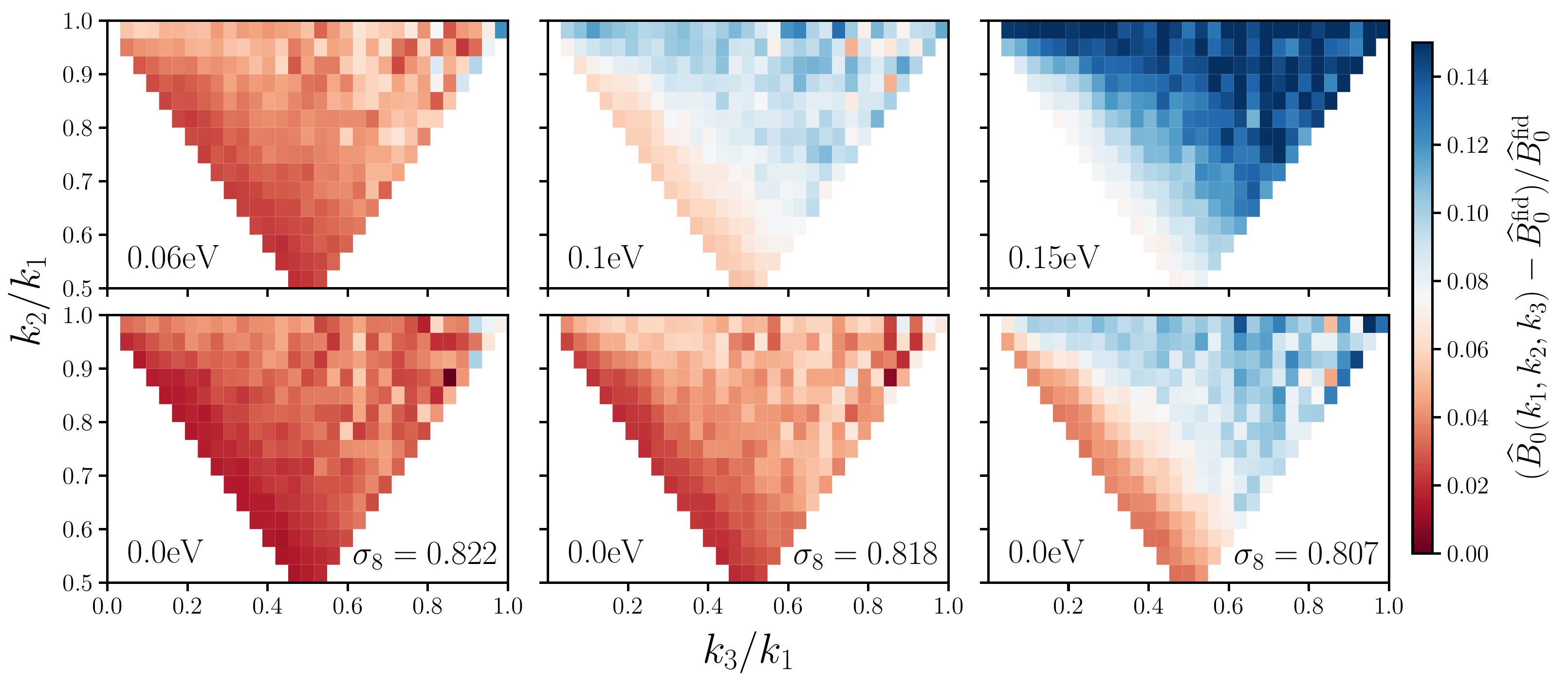} 
    \caption{The shape dependence of the $\smnu$ and $\sig$ imprint on 
    the redshift-space halo bispectrum, $\Delta \BOk/\BOk^\mathrm{fid}$. 
    We align the $\smnu{=}~0.06, 0.10$, and $0.15$ eV HADES simulations 
    in the upper panels with $\smnu{=}~0.0$~eV $\sig{=}~0.822$, $0.818$, and $0.807$ 
    simulations on the bottom such that the top and bottom panels in each column 
    have matching $\sig^{c}$, which produce mostly degenerate imprints on the 
    redshift-space power spectrum. $\BOk^{\rm fid}$ is measured from the 
    fiducial HADES simulation: $\smnu{=}~0.0$~eV and $\sig{=}~0.833$. 
    The difference between the top and bottom 
    panels highlight that $\smnu$ leaves a imprint distinct from $\sig$ on 
    elongated and isosceles triangles, bins along the bottom left and bottom 
    right edges, respectively. {\em The imprint of $\smnu$ has a distinct shape 
    dependence on the bispectrum that cannot be replicated by varying $\sig$}. 
    }
\label{fig:dbk_shape}
\end{center}
\end{figure}

\begin{figure}
\begin{center}
\includegraphics[width=0.9\textwidth]{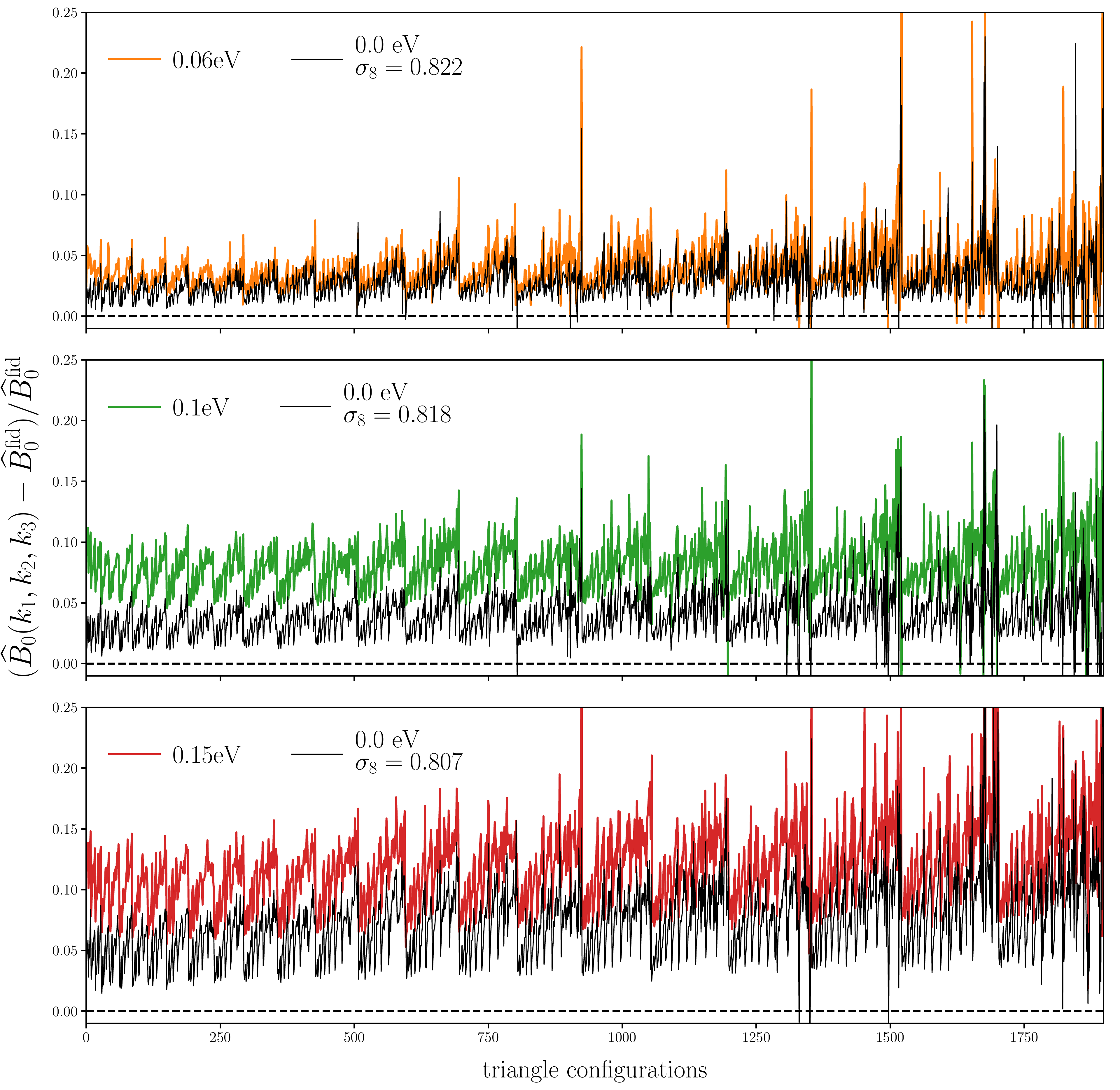}
    \caption{The impact of $\smnu$ and $\sig$ on the redshift-space halo bispectrum,
    $\Delta \BOk/\BOk^\mathrm{fid}$, for all $1898$ triangle configurations with 
    $k_1, k_2, k_3 \leq 0.5~\mpc$. We compare $\Delta \BOk/\BOk^\mathrm{fid}$ 
    of the $\smnu = 0.06$ (top), $0.10$ (middle), and $0.15$ eV (bottom) HADES simulations 
    to $\Delta \BOk/\BOk^\mathrm{fid}$ of $\smnu{=}~0.0$ eV, $\sig{=}~0.822$, $0.818$, and 
    $0.807$ simulations. The imprint of $\smnu$ on the bispectrum has a significantly different 
    amplitude than the imprint of $\sig$. For instance, $\smnu{=}~0.15\,\mathrm{eV}$ (red) 
    has a $\sim 5\%$ stronger impact on the bispectrum than $\smnu{=}~0.0$ eV  
    $\sig{=}~0.798$ (black) even though their power spectra only differ by $< 1\%$ 
    (Figure~\ref{fig:plk}). {\em The distinct imprint of $\smnu$ on the bispectrum 
    illustrate that the bispectrum can break the degeneracy between $\smnu$ and $\sig$ 
    that degrade constraints from two-point analyses}.
    }
\label{fig:dbk_amp}
\end{center}
\end{figure}
\subsection{Breaking the $\smnu$-- $\sig$ degeneracy} \label{sec:mnusig}
One major bottleneck of constraining $\smnu$ with the power spectrum alone is the 
strong $\smnu$ -- $\sig$ degeneracy. The imprints of $\smnu$ and $\sig$ on the power 
spectrum are degenerate and for models with the same $\sig^c$, the power spectrum
only differ by $< 1\%$ (see Figure~\ref{fig:plk} and~\citealt{villaescusa-navarro2018}). 
The HADES suite, which has simulations with $\smnu = 0.0, 0.06, 0.10$, 
and $0.15$ eV as well as $\smnu = 0.0$ eV simulations with matching $\sig^{c}$, 
provides an ideal set of simulations to disentangle the impact of $\smnu$ and 
examine the degeneracy between $\smnu$ and $\sig$ (Section~\ref{sec:hades} and Table~\ref{tab:sims}). 
We measure the bispectrum of the HADES simulations (Figure~\ref{fig:bk_shape} and~\ref{fig:bk_amp}) 
and present how the bispectrum can significantly improve $\smnu$ constraints 
by breaking the $\smnu$ -- $\sig$ degeneracy. 

We begin by examining the triangle shape dependent imprint of $\smnu$ on the 
redshift-space halo bispectrum versus $\sig$ alone. In Figure~\ref{fig:dbk_shape}, 
we present the fractional residual, $(\Delta \BOk = \BOk - \BOk^{\rm fid})/\BOk^{\rm fid}$,
as a function of $k_2/k_1$ and $k_3/k_1$ for $\smnu=0.06, 0.10$, and $0.15$~eV 
in the upper panels and 0.0~eV $\sig{=}~0.822$, $0.818$, and $0.807$ in the 
bottom panels. The simulations in the top and bottom panels of each column 
have matching $\sig^{c}$. Overall, as $\smnu$ increases, the amplitude of the 
bispectrum increases for all triangle shapes (top panels). This increase is due to halo 
bias~\citep[][see also Figure~\ref{fig:plk}]{villaescusa-navarro2018}. Since we 
impose a fixed $M_{\rm lim}$ on our halos, lower values of $\sig$ translate 
to a larger halo bias, which boosts the amplitude of the bispectrum. Within the 
overall increase in amplitude, however, there is a significant triangle dependence. 
Equilateral triangles (upper left) have the largest increase. For $\smnu=0.15$ eV, 
the bispectrum is $\sim 15\%$ higher than $\BOk^{\rm fid}$ for equilateral triangles 
while only $\sim 8\%$ higher for folded triangles (lower center). The noticeable 
difference in $\Delta \BOk/\BOk^{\rm fid}$ between equilateral and squeezed 
triangles (upper left) is roughly consistent with the comparison in Figure 7 
of~\cite{ruggeri2018}. They, however, fix $A_s$ in their simulations and 
measure the real-space halo bispectrum so we refrain from any detailed 
comparisons. 

As $\sig$ increases with $\smnu = 0.0$ eV, the bispectrum also increases 
overall for all triangle shapes (bottom panels). However, the comparison of the
top and bottom panels in each column reveals significant differences in 
$\Delta \BOk/\BOk^{\rm fid}$ for $\smnu$ versus $\sig$ alone. Between 
$\smnu=0.15$ eV and \{$0.0$ eV, $\sig = 0.807$\} cosmologies, there is an overall 
$\gtrsim 5\%$ difference in the bispectrum. In addition, the shape dependence 
of the $\Delta \BOk/\BOk^{\rm fid}$ increase is different for $\smnu$ than $\sig$. 
This is particularly clear in the differences between $0.1$ eV (top center panel) 
and $\{$0.0 eV, $\sig=0.807\}$ 
(bottom right panel): near equilateral triangles in the two panels have similar 
$\Delta \BOk/\BOk^{\rm fid}$ while triangle shapes near the lower left edge from 
the squeezed to folded triangles have significantly different $\Delta \BOk/\BOk^{\rm fid}$. 
Hence, $\smnu$ leaves an imprint on the bispectrum with a distinct triangle 
shape dependence than $\sig$ alone. In other words, the triangle shape dependent 
imprint of $\smnu$ on the bispectrum cannot be replicated by varying $\sig$ --- 
unlike the power spectrum. 

We next examine the amplitude of the $\smnu$ imprint on the redshift-space halo 
bispectrum versus $\sig$ alone as a function of all triangle configurations. We present 
$\Delta \BOk/\BOk^{\rm fid}$ for all $1898$ possible triangle configurations 
with $k_1, k_2, k_3 < k_{\rm max} = 0.5~h/{\rm Mpc}$ in Figure~\ref{fig:dbk_amp}. 
We compare $\Delta \BOk/\BOk^{\rm fid}$ of the $\smnu = 0.06, 0.10$, and 
$0.15$ eV HADES models to the $\Delta \BOk/\BOk^\mathrm{fid}$ of 
$\smnu{=}~0.0$ eV $\sig{=}~0.822$, $0.818$, and $0.807$ models in the
top, middle, and bottom panels, respectively. The comparison confirms the 
difference in overall amplitude of varying $\smnu$ and $\sig$ (Figure~\ref{fig:dbk_shape}). 
For instance, $\smnu{=}~0.15\,\mathrm{eV}$ (red) has a $\sim 5\%$ stronger 
impact on the bispectrum than $\smnu{=}~0.0$ eV $\sig{=}~0.798$ (black) 
even though their power spectra differ by $< 1\%$ (Figure~\ref{fig:plk}).

The comparison in the panels of Figure~\ref{fig:dbk_amp} confirms the difference 
in the configuration dependence in $\Delta \BOk/\BOk^\mathrm{fid}$ between $\smnu$ 
versus $\sig$. The triangle configurations are ordered by looping through $k_3$ 
in the inner most loop and $k_1$ in the outer most loop such that $k_1 \geq k_2 \geq k_3$. 
In this ordering, $k_1$ increases from left to right (see Figure~\ref{fig:bk_details} 
and Appendix~\ref{sec:bk_details}). $\Delta \BOk/\BOk^\mathrm{fid}$ 
of $\smnu$ expectedly increases with $k_1$: for small $k_1$ (on large scales), 
neutrinos behave like CDM and therefore the impact is reduced. However, 
$\Delta \BOk/\BOk^\mathrm{fid}$ of $\smnu$ has a smaller $k_1$ dependence than 
$\Delta \BOk/\BOk^\mathrm{fid}$ of $\sig$. The distinct imprint of $\smnu$ on the redshift-space 
halo bispectrum illustrates that the bispectrum can break the degeneracy between 
$\smnu$ and $\sig$. Therefore, by including the bispectrum, we can more precisely
constrain $\smnu$ than with the power spectrum. 

%% file: quijote.tex
\begin{figure}
\begin{center}
    \includegraphics[width=\textwidth]{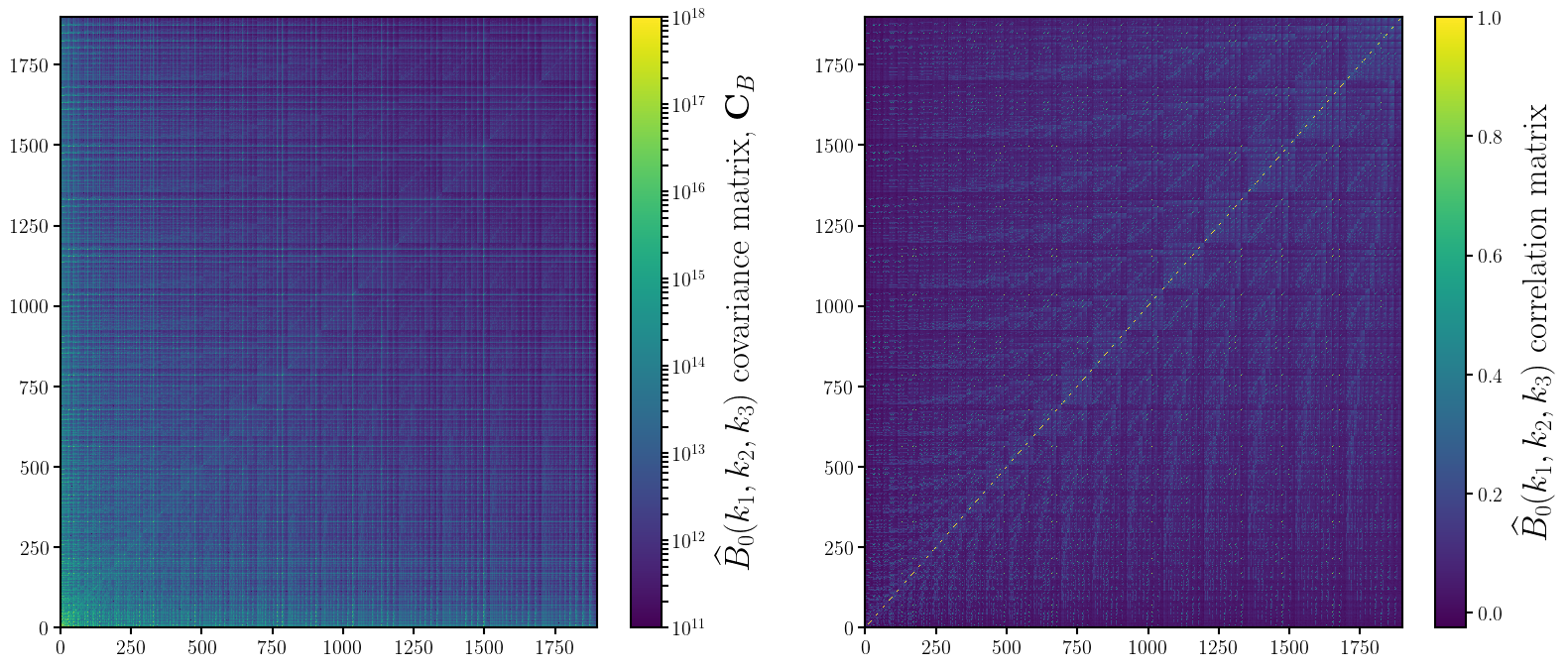} 
    \caption{Covariance and correlation matrices of the redshift-space halo bispectrum 
    estimated using $N_{\rm cov} = 15,000$ realizations of the Qujiote simulation suite at 
    the fiducial cosmology: $\Om{=}0.3175$, $\Ob{=}0.049$, $h{=}0.6711$, $n_s{=}0.9624$, $\sig{=}0.834$, 
    and $\smnu{=}0.0$ eV. We include all possible triangle configurations with 
    $k_1, k_2, k_3 \leq k_{\rm max} = 0.5~h/{\rm Mpc}$ and order the configurations 
    (bins) in the same way as Figures~\ref{fig:bk_amp} and~\ref{fig:dbk_amp}. We 
    use the covariance matrix above for the Fisher matrix forecasts presented in 
    Section~\ref{sec:forecasts}. 
    }
\label{fig:bk_cov}
\end{center}
\end{figure}

\begin{figure}
\begin{center}
    \includegraphics[width=0.8\textwidth]{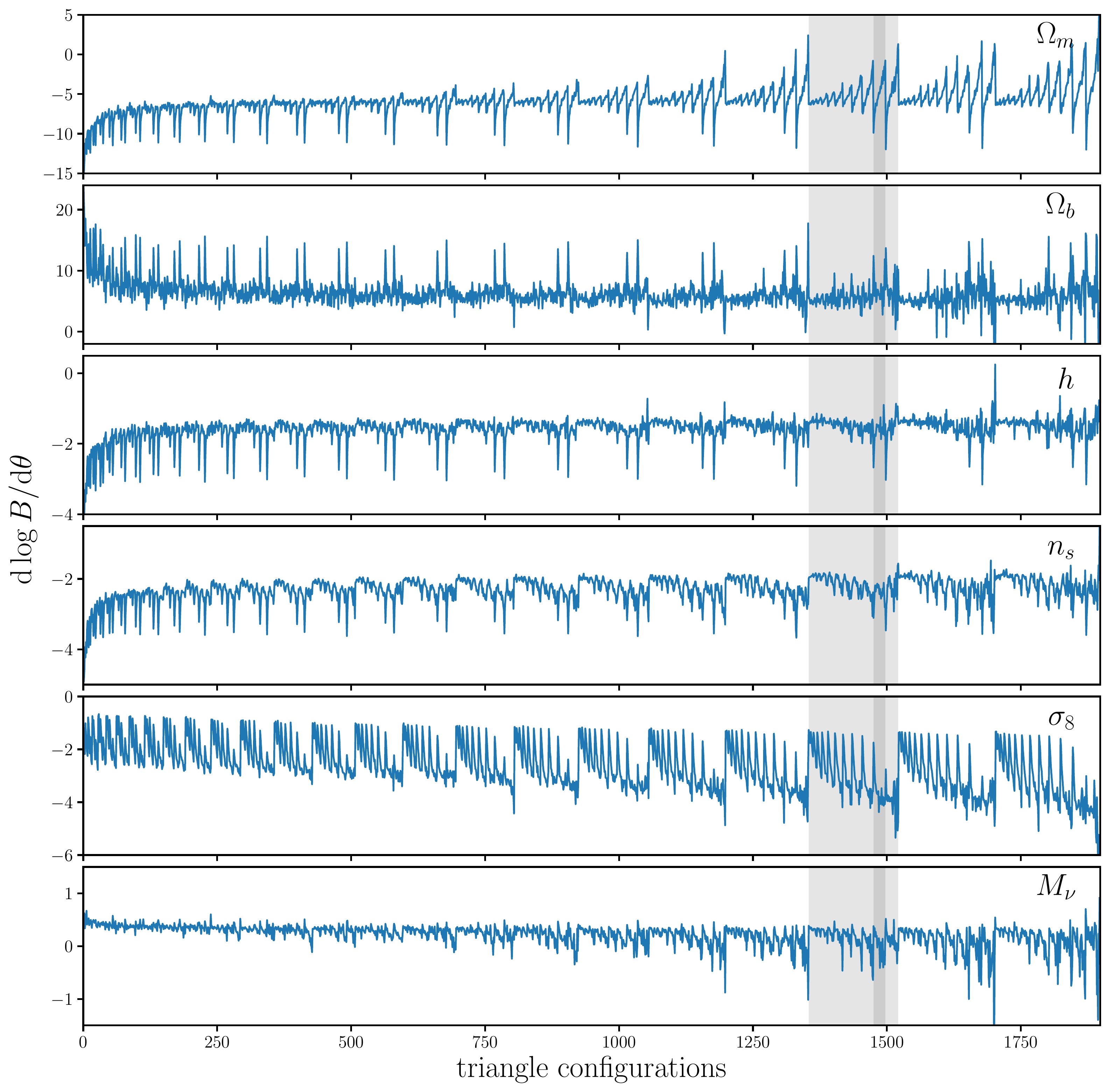} 
    \caption{Derivatives of the redshift-space halo bispectrum, ${\rm d}\log B_0 / {\rm d} \theta$,
    with respect to $\Om$, $\Ob$, $h$, $n_s$, $\sig$, and $\smnu$ as a function 
    of all $1898$ $k_1, k_2, k_3 \leq 0.5~\mpc$ triangles (top to bottom panels). 
    We estimate the derivatives at the fiducial parameters 
    using $N_{\rm deriv} = 1,500$ $N$-body realizations from the Quijote suite. 
    The configurations above are ordered in the same way as Figures~\ref{fig:bk_amp} 
    and~\ref{fig:dbk_amp} (Appendix~\ref{sec:bk_details}). The shaded region marks 
    configurations with $k_1 = 72k_f$; the darker shaded region marks configurations 
    with $k_1 = 72 k_f$ and $k_2 = 69 k_f$. 
    By using $N$-body simulations for the derivatives, we rely on fewer assumptions 
    and approximations than analytic methods (\emph{i.e.} perturbation theory). 
    \emph{Furthermore, with their accuracy on small scales, these derivatives enable 
    us to quantify for the first time the information content of the bispectrum in 
    the nonlinear regime.}
    }
\label{fig:bk_deriv}
\end{center}
\end{figure}

\subsection{$\smnu$ and other Cosmological Parameter Forecasts} \label{sec:forecasts}
We demonstrate in the previous section with the HADES simulations, that 
the bispectrum helps break the $\smnu$--$\sig$ degeneracy, a major 
challenge in precisely constraining $\smnu$ with the power spectrum. 
While this establishes the bispectrum as a promising probe for $\smnu$, 
we are ultimately interested in determining the constraining power of the 
bispectrum for an analysis that include cosmological parameters beyond 
$\smnu$ and $\sig$--- \emph{i.e.} $\Om$, $\Ob$, $h$, and $n_s$. The Quijote 
simulation suite is \emph{specifically} designed to answer this question
through Fisher matrix forecast.

First, the Quijote suite includes $N_{\rm cov} = 15,000$ $N$-body realizations run at a 
fiducial cosmology: $\smnu{=}0.0$eV, $\Om{=}0.3175, \Ob{=}0.049, 
n_s{=}0.9624, h{=}0.6711$, and $\sig{=}0.834$ (see Table~\ref{tab:sims}). 
This allows us to robustly estimate the $1898\times1898$ covariance matrix 
$\bfi{C}$ of the {\em full} bispectrum (Figure~\ref{fig:bk_cov}). Second, the 
Quijote suite includes $500$ $N$-body realizations evaluated at $13$ different 
cosmologies, each a small step away from the fiducial cosmology parameter 
values along one parameter (Section~\ref{sec:hades} and~Table~\ref{tab:sims}). 
We apply redshift-space distortions along 3 different directions for 
these $500$ realizations, which then effectively gives us $N_{\rm deriv.}{=}1,500$ 
realizations. These simulations allow us to precisely estimate the derivatives 
of the bispectrum with respect to each of the cosmological parameters. 

\begin{figure}
\begin{center}
    \includegraphics[width=\textwidth]{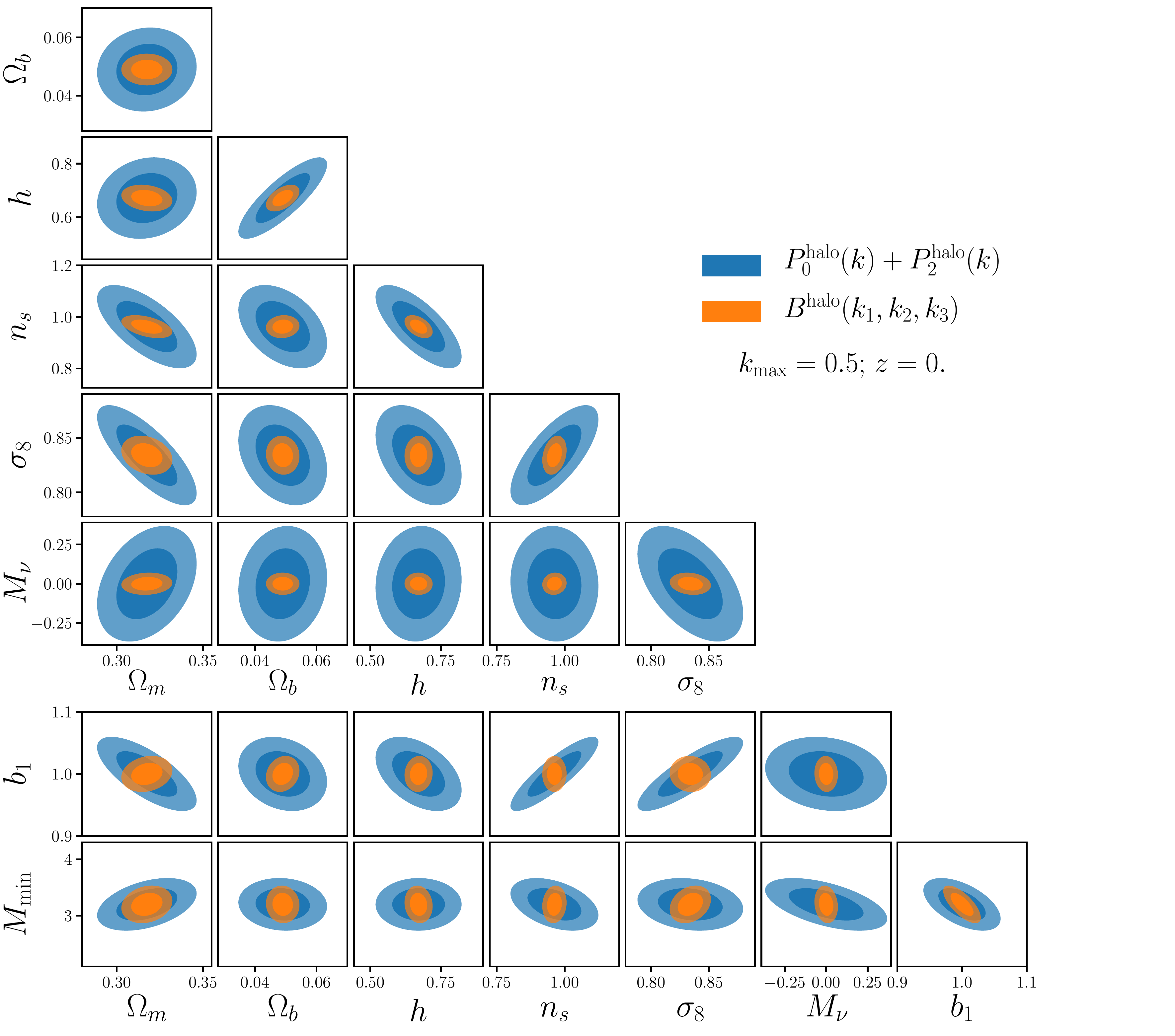} 
    \caption{Fisher matrix constraints for $\smnu$ and other cosmological parameters 
        for the redshift-space halo bispectrum monopole (orange). We include Fisher
        parameter constraints for the redshift-space halo power spectrum monopole and 
        quadrupole in blue for comparison. The contours mark the $68\%$ and $95\%$ 
        confidence intervals.  We set $k_{\rm max} = 0.5~h/{\rm Mpc}$ for both power 
        spectrum and bispectrum.  We include in our forecasts $b_1$ and $M_{\rm min}$, 
        linear bias (amplitude scaling factor) and halo mass limit, respectively. 
        They serve as a simplistic bias model and we marginalize over them so that 
        our constraints do not include extra 
        constraining power from the difference in bias/number density 
        in the different Quijote cosmologies. The bispectrum {\em substantially} 
        improves constraints on all of the cosmological parameters over the power 
        spectrum. Constraints on $\Om$, $\Ob$, $h$, $n_s$, and $\sig$ improve by
        factors of 1.9, 2.6, 3.1, 3.6, and 2.6, respectively. {\em For $\smnu$, the
        bispectrum improves $\sigma_{\smnu}$ from 0.2968 to 0.0572 eV --- over a
        factor of ${\sim}5$ improvement over the power spectrum}.}
\label{fig:bk_fish_05}
\end{center}
\end{figure}

Since their introduction to cosmology over two decades ago, Fisher information 
matrices have been ubiquitously used to forecast the constraining power of future 
experiments~\citep[\emph{e.g.}][]{jungman1996,tegmark1997,dodelson2003,heavens2009,verde2010}. 
Defined as 
\beq 
F_{ij} = - \bigg \langle \frac{\partial^2 \mathrm{ln} \mathcal{L}}{\partial \theta_i \partial \theta_j} \bigg \rangle,
\eeq
where $\mathcal{L}$ is the likelihood, the Fisher matrix for the bispectrum can 
be written as 
\beq \label{eq:fullfish} 
F_{ij} = \frac{1}{2}~\mathrm{Tr} \Bigg[\bfi{C}^{-1}\parti{\bfi{C}}\bfi{C}^{-1}\partj{\bfi{C}} + \bfi{C}^{-1} \left(\parti{B_0}\partj{B_0}^T + \parti{B_0}^T \partj{B_0} \right)\Bigg].
\eeq
Since we assume that the $B_0$ likelihood is Gaussian, including the first 
term in Eq.~\ref{eq:fullfish} runs the risk of incorrectly including information 
from the covariance already included in the mean~\citep{carron2013}. We, therefore,
conservatively neglect the first term and calculate the Fisher matrix as, 
\beq \label{eq:fisher}
F_{ij} = \frac{1}{2}~\mathrm{Tr} \Bigg[\bfi{C}^{-1} \left(\parti{B_0}\partj{B_0}^T + \parti{B_0}^T \partj{B_0} \right)\Bigg],
\eeq
directly with $\bfi{C}$ and $\partial B_0/\partial \theta_i$ along each cosmological 
parameter from the Quijote simulations. 
We include the Hartlap factor~\citep{hartlap2007} in our estimate of the 
precision matrix, $\bfi{C}^{-1}$, from $\bfi{C}$ to account for the bias caused 
from estimating $\bfi{C}^{-1}$ with a finite number of mocks.

For $\Om$, $\Ob$, $h$, $n_s$, and $\sig$, we estimate 
\beq \label{eq:dbkdt} 
\frac{\partial B_0}{\partial \theta_i} \approx \frac{B_0(\theta_i^{+})-B_0(\theta_i^{-})}{\theta_i^+ - \theta_i^-}, 
\eeq
where $B_0(\theta_i^{+})$ and $B_0(\theta_i^{-})$ are the average bispectrum of the 
$1,500$ realizations at $\theta_i^{+}$ and $\theta_i^{-}$, 
respectively. Meanwhile, for $\smnu$, where the fiducial value is 0.0 eV and we 
cannot have negative $\smnu$, we use the Quijote simulations at $\smnu^+$, 
$\smnu^{++}$, 
$\smnu^{+++}=0.1, 0.2, 0.4$ eV (Table~\ref{tab:sims}) to estimate 
\beq \label{eq:dbkdmnu} 
\frac{\partial B_0}{\partial \smnu} \approx \frac{-21 B_0(\theta_{\rm fid}^{\rm ZA}) + 
32 B_0(\smnu^{+}) - 12 B_0(\smnu^{++}) + B_0(\smnu^{+++})}{1.2}, 
\eeq
which provides a $\mathcal{O}(\delta \smnu^2)$ order approximation. 
Since the simulations at $\smnu^+$, $\smnu^{++}$, and $\smnu^{+++}$ are generated 
from Zel'dovich initial conditions, we use simulations at the fiducial cosmology 
also generated from Zel'dovich initial conditions ($\theta_{\rm fid}^{\rm ZA}$). 
By using these $N$-body simulations, instead of analytic methods (\emph{e.g.} perturbation theory), 
we exploit the accuracy of the simulations in the nonlinear regime and rely on fewer 
assumptions and approximations. In fact, these $N$-body simulation estimated 
derivatives are key in enabling us to quantify, for the first time, the total information 
content of the redshift-space bispectrum in the non-linear regime. We present the 
bispectrum derivatives for all triangle configurations in Figure~\ref{fig:bk_deriv} and  
discuss subtleties of the derivatives and tests of convergence and stability 
in Appendix~\ref{sec:numerical}.

We present the constraints on $\smnu$ and other cosmological parameters 
$\{\Om, \Ob, h, n_s, \sig\}$ derived from the redshift-space halo bispectrum 
Fisher matrix for $k_{\rm max} = 0.5~h/{\rm Mpc}$ in Figure~\ref{fig:bk_fish_05}. 
We include Fisher constraints for the redshift-space halo power spectrum 
monopole and quadrupole with the same $k_{\rm max}$ for comparison (blue). 
The shaded contours mark the $68\%$ and $95\%$ confidence intervals. 
We include in our Fisher constraints the following nuisance parameters: 
$b_1$ and $M_{\rm min}$. $b_1$ is the linear bias, a scaling factor on 
the bispectrum amplitude.
And $M_{\rm min}$ is the halo mass limit, which we choose as a nuisance
parameter to address the difference in the number densities among the Quijote
cosmologies, which impacts the derivatives $\partial B_0/\partial \theta_i$. 
For instance, the $\sig^{+}$ and $\sig^{-}$ cosmologies have halo 
$\bar{n} = 1.586\times10^{-4}$ and $1.528 \times 10^{-4}~h^{3}{\rm Mpc}^{-3}$. 
These parameters serve as a simplistic bias model and by marginalizing 
over them we aim to ensure that our Fisher constraints do not include extra 
constraining power from the difference in bias or number density. 
$b_1$ is a multiplicative factor so $\partial B_0/\partial b_1 \propto B_0$.
$\partial B_0/\partial M_{\rm min}$, we estimate numerically using 
$B_0$ evaluated at $M^{+}_{\rm min}{=}3.3\times10^{13}h^{-1}M_\odot$ 
and $M^{-}_{\rm min}{=}3.1\times10^{13}h^{-1}M_\odot$ with all other parameters 
set to the fiducial value. 
In the power spectrum constraints that we include for comparison, we also 
include $b_1$ and $M_{\rm min}$.

{\em The bispectrum substantially improves constraints on all parameters 
over the power spectrum.} For $k_{\rm max} = 0.5~h/{\rm Mpc}$, the 
bispectrum tightens the marginalized $1\sigma$ constraints, $\sigma_\theta$, of $\Om$, 
$\Ob$, $h$, $n_s$, and $\sig$ by factors of $\sim$1.9, 2.6, 3.1, 3.6, 2.6 over 
the power spectrum. {\em For $\smnu$, the bispectrum improves the constraint 
from $\sigma_{\smnu}{=}~0.2968$ to 0.0572 eV --- over a factor of 5 improvement 
over the power spectrum}. This $\sigma_{\smnu}{=}~0.0572$~eV constraint is from 
the {\em bispectrum alone} and only for a $1h^{-1}{\rm Gpc}$ box. For a larger 
volume, $V$, $\sigma_\theta$ scales roughly as $\propto1/\sqrt{V}$. 
The $\bar{n}$ of our halo catalogs (${\sim}1.56 \times 10^{-4}~h^3{\rm Mpc}^{-3}$) 
is also significantly lower than, for instance, $\bar{n}$ of the SDSS-III BOSS 
LOWZ + CMASS sample~\citep[${\sim}3 \times 10^{-4}~h^3{\rm Mpc}^{-3}$;][]{alam2015}. 
A higher number density reduces the shot noise term and thus would
further improve the constraining power of the bispectrum (see Eq.~\ref{eq:bk_sn}). 
We list the precise marginalized Fisher parameter constraints of both cosmological 
and nuisance parameters for $P_\ell$ and $B_0$ in Table~\ref{tab:forecast}. 
In this paper we focus on quantifying the information content and constraining 
power of $B_0$ alone. However, galaxy clustering studies typically analyze 
$P_\ell$ and $B_0$ jointly. We therefore also list the parameter constraints 
for a joint $P_\ell$ and $B_0$ analysis in Table~\ref{tab:forecast}.
For a joint analysis, $1\sigma$ constraints of $\{\Om, \Ob, h, n_s, \sig, \smnu\}$ 
tighten by factors of $\{\sim9.1, 8.8, 8.7, 9.5, 11.2, 8.1\times\}$ over a 
$P_\ell$ analysis and $\{\sim4.7, 3.3, 2.8, 2.6, 4.4, 1.6\times\}$ over a $B_0$ 
analysis.

\begin{figure}
\begin{center}
    \includegraphics[width=0.9\textwidth]{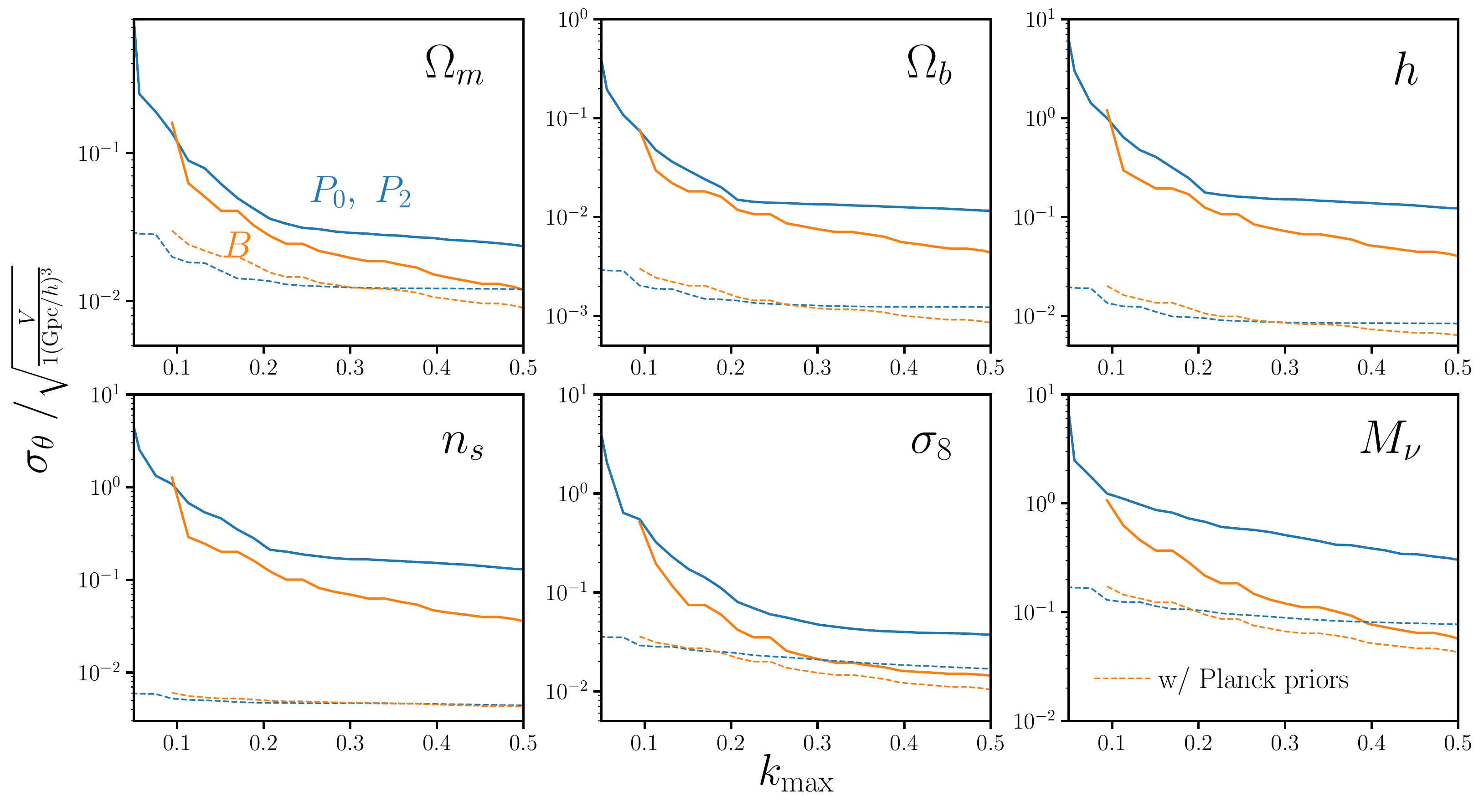} 
    \caption{Marginalized $1\sigma$ constraints, $\sigma_\theta$, of the cosmological 
    parameters $\Om$, $\Ob$, $h$, $n_s$, $\sig$, and $\smnu$ as a function 
    of $k_{\rm max}$ for the redshift-space halo bispectrum (orange) and power 
    spectrum ($\ell = 0, 2$; blue). The constraints are marginalized over the 
    nuisance parameters $b_1$ and $M_{\rm min}$ in our forecast (Section~\ref{sec:forecasts}).
    We impose $k_{\rm max} > 5~k_{\rm f}$ for $P_\ell$ and $k_{\rm max} > 12~k_{\rm f}$ 
    for $B_0$, the $k$ ranges where we have more data bins than number of parameters. 
    We also include $\sigma_\theta$ constraints with {\em Planck} priors (dotted). 
    Even below $k_{\rm max} < 0.5~h/{\rm Mpc}$, the bispectrum significantly 
    improves cosmological parameter constraints. The improvement, however, is larger 
    for higher $k_{\rm max}$. At $k_{\rm max} = 0.2~h/{\rm Mpc}$, the bispectrum 
    improves constraints on $\Om$, $\Ob$, $h$, $n_s$, $\sig$, and $\smnu$ by factors 
    of $\sim 1.3$, 1.1, 1.3, 1.5, 1.7, and 2.8 over the power spectrum. Even with 
    {\em Planck} priors, $B_0$ significantly improves $\sigma_\theta$ for $k_{\rm max} \gtrsim 0.2~\mpc$. 
    While the constraining power of $P_\ell$ saturates at $k_{\rm max} = 0.2~\mpc$,
    the constraining power of $B_0$ continues to increase out to $k_{\rm max} = 0.5~\mpc$.}
\label{fig:fish_kmax}
\end{center}
\end{figure}

Even below $k_{\rm max} < 0.5~h/{\rm Mpc}$, the bispectrum significantly 
improves cosmological parameter constraints. We compare $\sigma_\theta$ 
of $\Om$, $\Ob$, $h$, $n_s$, $\sig$, and $\smnu$ as a function of $k_{\rm max}$ 
for $B_0$ (orange) and $P_{\ell = 0,2}$ (blue) in Figure~\ref{fig:fish_kmax}. We only 
include the $k_{\rm max}$ range where the Fisher forecast is well defined --- 
\emph{i.e.} more data bins than the number of parameters: $k_{\rm max} > 5~k_{\rm f} \approx 0.03~\mpc$ 
for $P_\ell$ and $k_{\rm max} > 12~k_{\rm f} \approx 0.075~h/{\rm Mpc}$ for 
$B_0$. Figure~\ref{fig:fish_kmax} reveals that the improvement of the
bispectrum $\sigma_\theta$ over the power spectrum $\sigma_\theta$ is 
larger at higher $k_{\rm max}$. Although limited by the $k_{\rm max}$ range, 
the figure suggests that on large scales ($k_{\rm max}\lesssim 0.1~h/{\rm Mpc}$) 
$\sigma_\theta$ of $P_\ell$ crosses over $\sigma_\theta$ of $B_0$ so $P_\ell$ has more 
constraining power than $B_0$, as expected on linear scales. On slightly smaller 
scales, $k_{\rm max} = 0.2~h/{\rm Mpc}$, we find that the bispectrum improves 
$\sigma_\theta$ by factors of 
$\sim 1.3$, 1.1, 1.3, 1.5, 1.7, and 2.8 for $\Om$, $\Ob$, $h$, $n_s$, $\sig$, 
and $\smnu$ respectively. 

\begin{table}
    \caption{Marginalized Fisher parameter constraints from the redshift-space halo power 
    spectrum (top) and bispectrum (center) for different $k_{\rm max}$. At the bottom, we include 
    the parameter constraints from the joint redshift-space halopowerspectrum and bispectrum (bottom).
    We list constraints for cosmological parameters $\smnu$, $\Omega_m$, $\Omega_b$, $h$, $n_s$, 
    and $\sig$ as well as nuisance parameters $b_1$ and $M_{\rm min}$.} 
\begin{center} 
    \resizebox{\textwidth}{!}{\begin{tabular}{cccccccccc} \toprule
        & $k_{\rm max}$ & $\smnu$ & $\Omega_m$ & $\Omega_b$ & $h$ & $n_s$ & $\sig$ & $b_1$ & $M_{\rm min}$ \\
        & ({\footnotesize $h/{\rm Mpc}$}) &({\footnotesize eV}) & & & & & & & ({\footnotesize $10^{13} h^{-1}M_\odot$}) \\[3pt] \hline\hline

        &     & 0.0 & 0.3175 & 0.049 & 0.6711 & 0.9624 & 0.834 & 1. & 3.2  \\ 
$P_{\ell=0,2}$  
& 0.2 & $\pm0.678$ & $\pm0.036$ & $\pm0.015$ & $\pm0.177$ & $\pm0.211$ & $\pm0.080$ & $\pm0.105$ & $\pm1.660$\\
& 0.3 & $\pm0.510$ & $\pm0.029$ & $\pm0.013$ & $\pm0.151$ & $\pm0.167$ & $\pm0.047$ & $\pm0.058$ & $\pm0.725$\\
& 0.4 & $\pm0.390$ & $\pm0.027$ & $\pm0.013$ & $\pm0.139$ & $\pm0.153$ & $\pm0.040$ & $\pm0.051$ & $\pm0.482$\\
& 0.5 & $\pm{\bf 0.294}$ & $\pm0.023$ & $\pm0.012$ & $\pm0.122$ & $\pm0.130$ & $\pm0.037$ & $\pm0.048$ & $\pm0.375$\\
{\footnotesize +Planck priors} 
& 0.5 &$\pm0.077$& $\pm0.012$ & $\pm0.0012$ & $\pm0.0084$ & $\pm0.0044$ & $\pm0.017$ & $\pm0.016$ & $\pm0.212$\\\hline
$B_0$  
& 0.2 & $\pm0.217$ & $\pm0.028$ & $\pm0.012$ & $\pm0.124$ & $\pm0.124$ & $\pm0.042$ & $\pm0.079$ & $\pm1.135$\\
& 0.3 & $\pm0.119$ & $\pm0.020$ & $\pm0.008$ & $\pm0.072$ & $\pm0.069$ & $\pm0.021$ & $\pm0.043$ & $\pm0.587$\\
& 0.4 & $\pm0.078$ & $\pm0.015$ & $\pm0.006$ & $\pm0.052$ & $\pm0.047$ & $\pm0.016$ & $\pm0.029$ & $\pm0.367$\\
& 0.5 & $\pm{\bf 0.054}$ & $\pm0.011$ & $\pm0.004$ & $\pm0.039$ & $\pm0.034$ & $\pm0.014$ & $\pm0.023$ & $\pm0.254$\\
{\footnotesize +Planck priors} 
& 0.5 &$\pm0.043$& $\pm0.009$ & $\pm0.0009$ & $\pm0.0064$ & $\pm0.0043$ & $\pm0.010$ & $\pm0.023$ & $\pm0.253$ \\\hline
$P_{\ell=0,2}$ \& $B_0$
& 0.5 & $\pm{\bf 0.035}$ & $\pm0.002$ & $\pm0.0013$ & $\pm0.014$ & $\pm0.013$ & $\pm0.003$ & $\pm0.004$ & $\pm0.046$\\
{\footnotesize +Planck priors} & 0.5 &$\pm0.024$& $\pm0.002$ & $\pm0.0003$ & $\pm0.0025$ & $\pm0.0031$ & $\pm0.003$ & $\pm0.003$ & $\pm0.036$\\[3pt]




    \hline
\end{tabular}} \label{tab:forecast}
\end{center}
\end{table}

Our forecasts demonstrate that the bispectrum has significant constraining power 
beyond the power spectrum in the weakly nonlinear regime ($k > 0.1~\mpc$). 
This constraining power 
comes from the bispectrum breaking degeneracies among the cosmological and 
nuisance parameters. This is evident when we compare the unmarginalized 
constraints from $P_\ell$ and $B_0$: $1/\sqrt{F_{ii}}$ where $F_{ii}$ is a 
diagonal element of the Fisher matrix. For $k < 0.4~\mpc$, the unmarginalized 
constraints from $P_\ell$ are tighter than those from $B_0$. Yet, once we 
marginalize the constraints over the other parameters, the $P_\ell$ constraints 
are degraded and the $B_0$ constraints are tighter for $k > 0.1~\mpc$. The derivatives, 
$\partial B_0/\partial \theta_i$, also shed light on how $B_0$ breaks parameter 
degeneracies. The parameter degeneracies in the $P_\ell$ forecasts of Figure~\ref{fig:bk_fish_05} 
are consistent with similarities in the shape and scale dependence of 
$P_\ell$ derivatives $\partial P_0/\partial \theta$ and $\partial P_2/\partial \theta$. 

On the other hand, the $B_0$ derivatives with respect to the parameters 
have significant different scale and triangle shape dependencies.
In Figure~\ref{fig:bk_deriv}, we mark the triangles with $k_1 = 72 k_f$ in the 
shaded region and triangles with $k_1 = 72 k_f$ and $k_2 = 69 k_f$ in the darker 
shaded region. In the shaded region, $k_2$ increases for triangles to the right 
while $k_3$ and $\theta_{12}$, the angle between $k_1$ and $k_2$, increase for 
the triangles to the right in the darker shaded region. $\partial \log B_0/\partial \smnu$
has little scale dependence for triangles with small $\theta_{12}$ 
($k_1, k_2 \gg k_3$; folded or squeezed triangles). This is in contrast to the 
$\Om$, $\Ob$, $h$, and $n_s$ derivatives, which have significant scale dependence 
at $k < 0.2~\mpc$. For a given $k_1$ and $k_2$, $\partial \log B_0/\partial \smnu$
decreases as $\theta_{12}$ increases. The $\Om$, $h$, and $n_s$ derivatives have 
the opposite $\theta_{12}$ dependence. $\partial \log B_0/\partial \sig$
decreases as $\theta_{12}$ increases and also has little scale dependence for 
folded or squeezed triangles. However, the $\theta_{12}$ dependence of 
$\partial \log B_0/\partial \sig$, is scale independent unlike $\partial \log B_0/\partial \smnu$,
which has little $\theta_{12}$ dependence on large scales and stronger $\theta_{12}$ 
dependence on small scales. The detailed differences in the shape dependence of 
$\partial \log B_0/\partial \smnu$, $\partial \log B_0/\partial \sig$, other the 
other derivatives, highlighted in the shaded regions, ultimately allow $B_0$ 
to break parameter degeneracies. 

By exploiting the massive number of $N$-body simulations of the Quijote 
suite, we present for the first time the total information content of the 
full redshift-space bispectrum beyond the linear regime. The information content 
of the bispectrum has previously been examined using perturbation 
theory. Previous works, for instance, measure the signal-to-noise ratio (SNR) 
of the bispectrum derived from covariance matrices estimated using perturbation 
theory~\citep{sefusatti2005, sefusatti2006, chan2017}. More recently, \cite{chan2017},
used covariance matrices from 
$\sim 4700$ $N$-body simulations to find that the cumulative SNR of the halo bispectrum is 
$\sim 30\%$ of the SNR of the halo power spectrum at $k_{\rm max} \sim 0.1~\mpc$ 
and increases to $\sim 40\%$ at $k_{\rm max}\sim 0.35~\mpc$. While these simple 
SNR measurements cannot be easily compared to Fisher analysis~\citep{repp2015, blot2016}, 
we note that they are roughly consistent with the unmarginalized constraints, 
which loosely represent the SNRs of the derivatives. 
Also, when we measure the the halo power spectrum and bispectrum SNRs using our 
covariance matrices (Figure~\ref{fig:bk_cov}), we find a relation between the 
SNRs consistent with \cite{chan2017}. Beyond the $k$ range explored by \cite{chan2017},  
$k_{\rm max}>0.35~\mpc$, we find that the SNR of $B_0$ continues to increase 
at higher $k_{\rm max}$ in contrast to the $P_0$ SNR, which saturates at 
$k_{\rm max}\sim 0.1~\mpc$. At $k_{\rm max} = 0.75~\mpc$, the largest $k$ we 
measure the $B_0$, the SNR of $B_0$ is $\sim75\%$ of the SNR of $P_0$.

Beyond these signal-to-noise calculations, a number of previous works 
have quantified the information content of the bispectrum~\citep{scoccimarro2004, sefusatti2006, sefusatti2007, song2015, tellarini2016, yamauchi2017a, karagiannis2018, yankelevich2019, chudaykin2019, coulton2019, reischke2019}. 
While most of these works fix most cosmological parameters and focus solely 
on forecasting constraints of primordial non-Gaussianity and bias parameters, 
\cite{sefusatti2006}, \cite{yankelevich2019}, and \cite{chudaykin2019} provide 
bispectrum forecasts for full sets of cosmological parameters. In \cite{sefusatti2006}, 
they present likelihood analysis forecasts for $\omega_d$, $\omega_b$, $\OL$, 
$n_s$, $A_s$, $w$, $\tau$. 
For $\Lambda$CDM, with fixed bias parameters, and $k_{\rm max} = 0.3~\mpc$, 
they find constraints on $\Om$, $\Ob$, $h$, $n_s$, and $\sig$ from WMAP, 
$P_0$, and $B_0$ is $\sim 1.5$ times tighter than constraints from WMAP 
and $P_0$. In comparison, for 
$k_{\rm max} = 0.3~\mpc$ our $B_0$, constraints are tighter than $P_0$ constraints 
by factors of 1.4, 1.7, 2.0, 2.3, 2.0, and 3.9. Both \cite{sefusatti2006} and our 
analysis find significantly tighter constraints with the bispectrum. 
They however include the WMAP likelihood in their forecast and use 
perturbation theory models, which is limited to larger scales than used 
here~\citep{scoccimarro1998a, scoccimarro1999a, sefusatti2010, pollack2012, gil-marin2014, lazanu2016, eggemeier2019}.

\cite{yankelevich2019} present Fisher forecasts for $\Omega_{\rm cdm}$, 
$\Ob$, $h$, $n_s$, $A_s$, $w_0$, and $w_0$ for a 
Euclid-like survey~\citep{laureijs2011} in 14 non-overlapping redshift 
bins over $0.65 < z < 2.05$. They use the full redshift-space bispectrum,
rather than just the monopole, and a more sophisticated bias expansion 
than \cite{sefusatti2006} but use a perturbation theory bispectrum model, 
which consequently limit their forecast to $k_{\rm max} = 0.15~\mpc$. 
They find similar constraining power on cosmological parameters from $B$ 
alone as $P$. They also find that combining the bispectrum with the power 
spectrum only moderately improves parameter constraints because posterior 
correlations are similar for $P$ and $B$. While this seemingly conflicts 
with the results we present, there are significant differences between our 
forecasts. For instance, they forecasts the Euclid survey 
(\emph{i.e.} $z > 0.7$), while our forecasts are for $z = 0$.
They also forecast the {\em galaxy} $P$ and $B$ and marginalizes over 
56 nuisance parameters (14 $z$ bins each with 3 bias parameters and 1 
RSD parameter). They also neglect non-Gaussian contributions to the $B$ 
covariance matrix, which play a significant role on small scales~\citep{chan2017} 
and may impact the constraints.  
Despite differences, \cite{yankelevich2019} find that the constraining 
power of $B$ relative to $P_\ell$ increases for higher $k_{\rm max}$, consistent 
with our forecasts as a function of $k_{\rm max}$ (Figure~\ref{fig:fish_kmax}). 
Also, consistent with their results, for $k_{\rm max} = 0.15~\mpc$, 
we find similar posterior correlations between the $P_\ell$ and $B_0$ 
constraints. At $k_{\rm max} = 0.5~\mpc$, however, we find the posterior 
correlations are no longer similar, which contribute to the constraining
power of $B_0$~(Figure~\ref{fig:bk_fish_05}).

Finally, \cite{chudaykin2019} present power spectrum and bispectrum 
forecasts for $\omega_{\rm cdm}$, $\omega_b$, $h$, $n_s$, $A_s$, $n_s$, 
and $\smnu$ of a Euclid-like survey in 8 non-overlapping redshift bins 
over $0.5 < z < 2.1$. They use
a one-loop perturbation theory model for the redshift-space  power spectrum
multipoles ($\ell = 0, 2, 4$) and a tree-level bispectrum monopole model. 
Also, rather than imposing a $k_{\rm max}$ cutoff to restrict their 
forecasts to scales where their perturbation theory model can be trusted, 
they use a theoretical error covariance model approach from~\cite{baldauf2016}.
They find ${\sim}1.4$, 1.5, 1.2, 1.5, and 1.3 times tighter constraints 
from $P_\ell$ and $B_0$ than from $P_\ell$ alone. For $\smnu$, they find 
a factor of 1.4 improvement, from 0.038 eV to 0.028 eV. Overall, 
\cite{chudaykin2019} find more modest improvements from including $B_0$ than 
the improvements we find in our $k_{\rm max} = 0.5~\mpc$ $B_0$ constraints 
over the $P_\ell$ constraints. 

Among the differences from our forecast, most notably, \cite{chudaykin2019}  
marginalize over 64 parameters of their bias model (4 bias and 4 counterterm 
normalization parameters at each redshift bin). They also include the 
Alcock-Paczynski (AP) effect for $P_\ell$, which significantly improve the 
$P_\ell$ parameter constraints (\emph{e.g.} tightens $\smnu$ constraints by $\sim30\%$). 
They however, do not include AP effects in $B_0$. Furthermore, 
\cite{chudaykin2019} use theoretical error covariance to quantify the 
uncertainty of their perturbation theory models. Because they use the tree-level 
perturbation theory model for $B_0$, theoretical errors for $B_0$ quickly 
dominate at $k \gtrsim 0.1~\mpc$, where the one- and two-loop contribute 
significantly~\citep[\emph{e.g.}][]{lazanu2018}. Hence, their forecasts do not include 
the constraining power on nonlinear scales. Also different from our analysis,
\cite{chudaykin2019} use an Markov-Chain Monte-Carlo (MCMC) 
approach, which derives more accurate parameter constraints than our Fisher
approach. They, however, neglect non-Gaussian contributions to both the 
$P_\ell$ and $B_0$ covariance matrices and do not include the covariance 
between $P_\ell$ and $B_0$ for their joint constraints. While their theoretical 
error covariance, which couples different $k$-modes, may partly take this into 
account, we find that neglecting the covariance between $P_\ell$ and $B_0$ overestimates 
constraints, tighter by ${\sim}20\%$ for $k_{\rm max} = 0.2~\mpc$. 
Nonetheless, \cite{chudaykin2019} 
and our results both find significant improves in cosmological parameter 
constraints from including $B_0$. In fact, taking the theoretical errors 
into account, the improvements from $B_0$ they find are loosely consistent 
with our results at $k_{\rm max} \sim  0.2~\mpc$. 

Various differences between our forecast and previous work prevent more 
thorough comparisons. However, crucial aspects of our simulation based 
approach distinguish our forecasts from other works. 
We present the first bispectrum forecasts for a full set of cosmological 
parameters using bispectrum measured entirely from $N$-body simulations.
By using the simulations, we go beyond perturbation theory models and 
accurately model the redshift-space bispectrum to the nonlinear regime. 
Furthermore, by exploiting the immense number of simulations, we 
accurately estimate the full high-dimensional covariance matrix of 
the bispectrum. With these advantages, we present the first forecast
of cosmological parameters from the bispectrum down to nonlinear scales
and demonstrate the constraining power of the bispectrum for $\smnu$. 
Below, we underline a few caveats of our forecasts.

Our forecasts are derived from Fisher matrices. Such forecasts make 
the assumption that the posterior is approximately Gaussian and, as a result, 
they underestimate the constraints for posteriors that are highly 
non-elliptical or asymmetric~\citep{wolz2012}. Fisher matrices also rely 
on the stability, and in our case also convergence, of numerical derivatives. 
We examine the stability of the $B_0$ derivatives with respect to $\smnu$ 
by comparing the derivatives computed using $N$-body simulations 
at three different sets of cosmologies: (1) \{$\theta_{\rm fid}^{\rm ZA}$, $\smnu^{+}$, $\smnu^{++}$, 
$\smnu^{+++}$\} (Eq.~\ref{eq:dbkdmnu}), (2) \{$\theta_{\rm fid}^{\rm ZA}$, $\smnu^{+}$, 
$\smnu^{++}$\}, and (3) \{$\theta_{\rm fid}^{\rm ZA}$, $\smnu^{+}$ \}~(see Appendix~\ref{sec:numerical}; Figure~\ref{fig:dPBdmnu}). 
The derivatives computed using the different set of cosmologies, do not impact 
the $\Om$, $\Ob$, $h$, $n_s$, and $\sig$ constraints. They do however affect the 
$\smnu$ constraints; but because $P_\ell$ and $B_0$ derivatives are affected by 
the same factor, the relative improvement of the $B_0$ $\smnu$ constraint over 
the $P_\ell$ constraints is \emph{not} impacted. 
In addition to the stability, because we use $N$-body simulations, we test 
whether the convergence of our covariance matrix and derivatives impact 
our forecasts by varying the number of simulations 
used to estimate them: $N_{\rm cov}$ and $N_{\rm deriv}$, respectively. 
For $N_{\rm cov}$, we find $< 5\%$ variation in the Fisher matrix elements, 
$F_{ij}$, for $N_{\rm cov} > 5000$ and $< 1\%$ variation in $\sigma_\theta$
for $N_{\rm cov} > 12000$.  For $N_{\rm deriv}$, we find $< 5\%$ variation 
in the $F_{ij}$ elements and $< 5\%$ variation in $\sigma_\theta$ for 
$N_{\rm deriv} > 1200$. Since our constraints vary by $< 10\%$ for sufficient 
$N_{\rm cov}$ and $N_{\rm deriv}$, the convergence of the covariance matrix 
and derivatives do not impact our forecasts to the accuracy level of Fisher 
forecasting. We refer readers to Appendix~\ref{sec:numerical} for a more 
details on the robustness of our results to the stability of the derivatives 
and convergence. 

We argue that the constraining power of the bispectrum and its improvement 
over the power spectrum come from breaking degeneracies among the cosmological 
parameters. However, numerical noise can impact our forecasts when we invert 
the Fisher matrix. Since {\em Planck} constrain $\{\Om$, $\Ob$, $h$, $n_s$, $\sig\}$ 
tighter than either $P_\ell$ or $B_0$ alone, the elements of the {\em Planck} prior 
matrix are larger than the elements of $P_\ell$ and $B_0$ Fisher matrices. 
Including {\em Planck} priors (\emph{i.e.} adding the prior matrix to the 
Fisher matrix) increases the numerical stability of the matrix inversion. 
It also reveals whether the bispectrum still improves parameter constraints 
once we include CMB constraints. With {\em Planck} priors and $P_\ell$ to 
$k_{\rm max} = 0.5~\mpc$, we derive the following constraints: 
$\sigma_{\Om} = 0.0120$, $\sigma_{\Ob} = 0.0012$, $\sigma_h=0.0084$, 
$\sigma_{n_s}=0.0044$, and $\sigma_{\sig}=0.0169$. 
Including {\em Planck} priors expectedly tighten the constraints from $P_\ell$. 
Meanwhile, with {\em Planck} priors and $B_0$ to $k_{\rm max} = 0.5~\mpc$, we get 
$\sigma_{\Om} = 0.0090$, $\sigma_{\Ob} = 0.0009$, $\sigma_h=0.0064$, 
$\sigma_{n_s}=0.0043$, and $\sigma_{\sig}=0.0104$, 1.3, 1.4, 1.3, 1.0, and 1.6
times tighter constraints. For $\smnu$, $\sigma_{\smnu} = 0.0773$ eV for $P_\ell$ 
and $\sigma_{\smnu} = 0.0430$ eV for $B_0$, a factor of 1.8 improvement. Since we find
substantial improvements in parameter constraints with the {\em Planck} prior, 
the improvement from $B_0$ are numerically robust. Furthermore, $\sigma_\theta$ 
as a function of $k_{\rm max}$ with {\em Planck} priors reveal that while the 
constraining power of $P_\ell$ saturates at $k_{\rm max} = 0.2~\mpc$, the constraining 
power of $B_0$ continues to increase out to $k_{\rm max} = 0.5~\mpc$ 
(dotted; Figure~\ref{fig:fish_kmax}). 

Our forecasts are derived using the power spectrum and bispectrum in
\emph{periodic boxes}. We do not consider a realistic geometry or radial selection 
function of actual observations from galaxy 
surveys. A realistic selection function will smooth out the triangle 
configuration dependence and consequently degrade the constraining power 
of the bispectrum. In \cite{sefusatti2005}, for instance, they find that the 
signal-to-noise of the bispectrum is significantly reduced once survey geometry 
is included in their forecast. We also do not account for super-sample 
covariance~\citep[\emph{e.g.}][]{hamilton2006, sefusatti2006, takada2013, li2018}, 
which may also degrade constraints. Survey geometry and super-sample covariance, 
however, also degrades the signal-to-noise of their power spectrum forecasts. 
Hence, with the substantial improvement in the $\smnu$ constraints of the bispectrum, 
even with survey geometry we expect the bispectrum will significantly improve 
$\smnu$ constraints over the power spectrum.  

We include the nuisance parameter $\mmin$ in our forecasts to address the 
difference in halo bias and number densities among the Quijote cosmologies. Although 
we marginalize over $\mmin$, this may not fully account for the extra 
information from $\bar{n}$ and nonlinear bias leaking into the derivatives. 
To test this, we include extra nuisance parameters, $\{A_{\rm SN}$, 
$B_{\rm SN}$, $b_2$, $\gamma_2\}$, and examine their impact on our forecasts. 
$A_{\rm SN}$ and $B_{\rm SN}$ account for any $\bar{n}$ dependence that may 
be introduced from the shot noise correction. They are multiplicative factors 
of the first and second terms of Eq.~\ref{eq:bk_sn}. $b_2$ and $\gamma_2$ 
are the quadratic bias and nonlocal bias parameters~\citep{chan2012, sheth2013}
to account for information from nonlinear bias.
Marginalizing over $\{b_1$, $\mmin$, $A_{\rm SN}$, $B_{\rm SN}$, $b_2$, $\gamma_2\}$, 
we obtain the following constraints for $B_0$ with $k_{\rm max} = 0.5~\mpc$:
$\sigma_{\Om} = 0.0129$, $\sigma_{\Ob} = 0.0044$, $\sigma_h=0.0404$, 
$\sigma_{n_s}=0.0456$, $\sigma_{\sig}=0.0228$, and $\sigma_{\smnu}=0.0575$.  
While constraints on $n_s$ and $\sig$ are broadened from our fiducial 
forecasts, by $27\%$ and $60\%$, the other parameters, especially $\smnu$, 
are not significantly impacted by marginalizing over the extra nuisance 
parameters. As another test, we calculate derivatives using halo catalogs 
from Quijote $\theta^{-}$ and $\theta^{+}$ cosmologies with fixed $\bar{n}$. 
We similarly find no significant impact on the $B_0$ parameter constraints.
Forecasts using additional nuisance parameters and with fixed $\bar{n}$ 
derivatives, both support the robustness of our forecast. Yet these 
tests do not ensure that our forecast entirely marginalizes over halo bias. 

In this paper, we focus on the halo bispectrum and power spectrum. However,   
constraints on $\smnu$ will ultimately be derived from the distribution of 
galaxies. Besides the cosmological parameters, bias and nuisance parameters 
that allow us to marginalize over galaxy bias need to be incorporated to 
forecast $\smnu$ and other cosmological parameter constraints for the 
galaxy bispectrum. 
Although we include a \emph{naive} bias model through $b_1$ and 
$M_{\rm min}$, and even $b_2$ and $\gamma_2$ in our tests, this is 
insufficient to describe how galaxies trace matter. 
A more realistic bias model 
such as a halo occupation distribution (HOD) model involve extra parameters 
that describe the distribution of central and satellite galaxies in 
halos~\citep[\emph{e.g.}][]{zheng2005,leauthaud2012,tinker2013,zentner2016,vakili2019}. 
We, therefore, refrain from a more exhaustive investigation of the impact of 
halo bias on our results and focus quantifying the constraining power of 
the galaxy bispectrum in the next paper of this series: Hahn et al. (in preparation). 

Marginalizing over galaxy bias parameters, will likely reduce the constraining 
power at high $k$. 
Improvements that come from extending to smaller scales will also be diminished 
by the extra parameters needed to accurately model those scales. \cite{hand2017}, 
for instance, using a 13 parameter model only find a 15-30\% improvement in 
$f\sig$ when they extend their power spectrum multipole analysis from $0.2~\mpc$
to $0.4~\mpc$. Although we focus on parameter constraints from the bispectrum 
alone in this work, jointly analyzing the power spectrum and bispectrum will 
improve matters in this regard. A combined analysis will help constrain bias
parameters, further break parameter degeneracies, and improve constraints on 
cosmological parameters~\citep{sefusatti2006, yankelevich2019, chudaykin2019, coulton2019}.
Already for our bias model, we found substantial improvements in the
cosmological parameter constraints forecasts for a joint $P_\ell$ and $B_0$ analysis
(Table~\ref{tab:forecast}).
Furthermore, we emphasize that the constraints we present is for a $1h^{-1}{\rm Gpc}$ 
box and $\bar{n} \sim 1.56 \times 10^{-4}~h^3{\rm Mpc}^{-3}$, a substantially 
smaller volume and lower number density than upcoming surveys. 
Thus, even if the constraining 
power at high $k$ is reduced, our forecasts suggest that the bispectrum offers 
significant improvements over the power spectrum, especially for constraining $\smnu$.

%% file: summary.tex
\section{Summary} 
A precise measurement of $\smnu$ can distinguish between the `normal' 
and `inverted' neutrino mass hierarchy scenarios and reveal physics 
beyond the Standard Model of particle physics. The total neutrino mass, 
through impact the  expansion history and the growth of cosmic structure 
in the Universe, can be measured using cosmological observables (\emph{e.g.} 
CMB and large-scale structure). In fact, cosmological probes have the 
potential to more precisely measure $\smnu$ than current and upcoming 
laboratory experiments. The current tightest cosmological $\smnu$ 
constraints come from combining CMB data with other cosmological probes. 
The degeneracy between $\smnu$ and $\tau$, the optical depth of reionization, 
however, is a major bottleneck for $\smnu$ constraints from CMB data 
that will {\em not} be addressed by upcoming ground-based CMB experiments. 
Measuring the imprint of neutrinos on the 3D clustering of galaxies 
provides a promising alternative to tightly constrain $\smnu$, especially 
with the unprecedented cosmic volumes that will be mapped by upcoming 
surveys. 

Recent developments in simulation based emulation methods have improved 
the accuracy of theoretical predictions beyond linear scales and address 
the challenges of unlocking the information content in nonlinear clustering 
to constrain $\smnu$. Yet, for power spectrum analyses, the strong 
degeneracy between the imprints of $\smnu$ and $\sig$ poses a serious 
limitation for constraining $\smnu$. Information in the nonlinear 
regime, however, cascade from the power spectrum to higher-order statistics 
such as the bispectrum. Previous works have demonstrated that the bispectrum
has comparable signal-to-noise as the power spectrum on nonlinear scales~\citep{sefusatti2005, chan2017}
and that it improves constraints on cosmological parameters~\citep{sefusatti2006, yankelevich2019, chudaykin2019}. 
No work to date has quantified the total information content and constraining 
power of the full bispectrum down to nonlinear scales. 

In this work, we examined the effect of massive neutrinos on the redshift-space 
halo bispectrum using ${\sim}23,000$ $N$-body simulations with massive neutrinos
from the HADES and Quijote suite. Using $N$-body simulations and with such a massive 
number of them, we directly address key challenges of accurately modeling the 
bispectrum in the nonlinear regime and estimating its high dimensional covariance 
matrix. More specifically, 
\begin{itemize}
    \item We examine the imprint of $\smnu$ and $\sig$ on the redshift-space halo bispectrum, 
        $B_0$, using two sets of HADES simulations. One with massive neutrinos, $\smnu = 0.0$, 0.06, 
        0.1, and 0.15 eV; the other with $\smnu = 0.0$ eV and matching $\sig^{c}$. 
        $\smnu$ and $\sig$ leave distinct imprints on $B_0$ with significantly different 
        scale and triangle shape dependencies. Thus, we demonstrate that $B_0$ helps 
        break the $\smnu$-$\sig$ degeneracy found in the power spectrum. 
    \item We quantify the total information content of $B_0$ using a Fisher matrix forecast of 
        $\{\Om$, $\Ob$, $h$, $n_s$, $\sig$, $\smnu\}$. With bispectrum measured from 22,000 
        $N$-body simulations from the Quijote suite, we are able to derive stable and converged 
        covariance matrix and derivatives for our forecast. Furthermore, our simulation 
        based approach allows us to extend beyond perturbation theory and use all 1898 
        triangle configurations of the bispectrum down to $k_{\rm max} = 0.5~\mpc$.
    \item For $k_{\rm max}{=}0.5~\mpc$, the bispectrum produces $\Om$, $\Ob$, $h$, $n_s$, and 
        $\sig$ constraints 1.9, 2.6, 3.1, 3.6, and 2.6 times tighter than the power spectrum. 
        For $\smnu$, we derive 1$\sigma$ constraint of 0.0572 eV --- over 5 times tighter than 
        the power spectrum. Even with priors from {\em Planck}, the bispectrum improves 
        $\smnu$ constraints by a factor of 1.8. These constraints are derived for a 
        $(1~h^{-1}{\rm Gpc})^3$ box, a substantially smaller volume than upcoming surveys.
\end{itemize}

While our results clearly showcase the advantages of the bispectrum for
more precisely constraining $\smnu$, as well as the cosmological parameters, 
a number of assumptions go into our forecast. Fisher matrix forecasts assume 
that the posterior is approximately Gaussian and, as a result, they 
overestimate the constraints for posteriors that are highly non-elliptical 
or asymmetric. Furthermore, our forecasts are derived using the power spectrum 
and bispectrum in periodic boxes, instead of a realistic geometry or 
radial selection function of galaxy surveys, which degrades the constraining power. 
Since we use periodic boundary conditions, we also do not account for super-sample
covariance, which in practice results from a non-trivial survey geometry. Lastly, 
our forecast focuses on the halo bispectrum, while the \emph{galaxy} bispectrum 
is what is actually measured from observations. 
We marginalize over a simplistic bias model through $b_1$ and $M_{\rm min}$; 
however, such a model is insufficient to 
flexibly describe how galaxies trace matter. A more realistic bias model 
involve extra parameters that describe the distribution of central and satellite 
galaxies. In the next paper of the series, we include HOD parameters in our
forecasts and quantify the full information content of the redshift-space 
galaxy bispectrum. 

The constraints from our forecasts are derived for a $(1~h^{-1}{\rm Gpc})^3$ 
volume. These constraints roughly scale as $\propto1/\sqrt{V}$ with volume. 
Upcoming spectroscopic galaxy surveys will map out vastly larger cosmic
volumes: PFS $\sim 9~h^{-3}{\rm Gpc}^3$ and DESI $\sim 50~h^{-3}{\rm Gpc}^3$~\citep{takada2014, desicollaboration2016}. 
Euclid and WFIRST, space-based surveys, will expand these volumes to 
higher redshifts. A key science goal in these surveys will be constraining $\smnu$. 
Scaling our 0.0572 eV $\smnu$ constraint from the bispectrum to the survey 
volumes gives 0.0191 eV and 0.0081 eV $1\sigma$ constraints for PFS and DESI. 
Galaxy samples in these surveys will also have significantly higher 
number densities than the halos used in our forecast: \emph{e.g.} 
the DESI BGS at $z{\sim}0.3$, DESI LRG, and PFS at $z \sim 1.3$ 
will have ${\sim}20, 3$, and $5\times$ higher number densities, 
respectively. With such precision, we would detect $\smnu$ by $>3\sigma$ and 
distinguish between the normal and inverted neutrino mass hierarchy scenarios 
by $>2\sigma$ from the bispectrum alone. Such naive projections, however, 
should be taken with more than a grain of salt. Nevertheless, the substantial 
improvement we find in constraints with the bispectrum over the power spectrum, 
strongly advocate for analyzing future surveys with more than the power spectrum. 
Our results demonstrate the potential of the bispectrum to tightly constrain 
$\smnu$ with unprecedented precision.

%% file: rsd.tex
\begin{figure}
\begin{center}
    \includegraphics[width=0.95\textwidth]{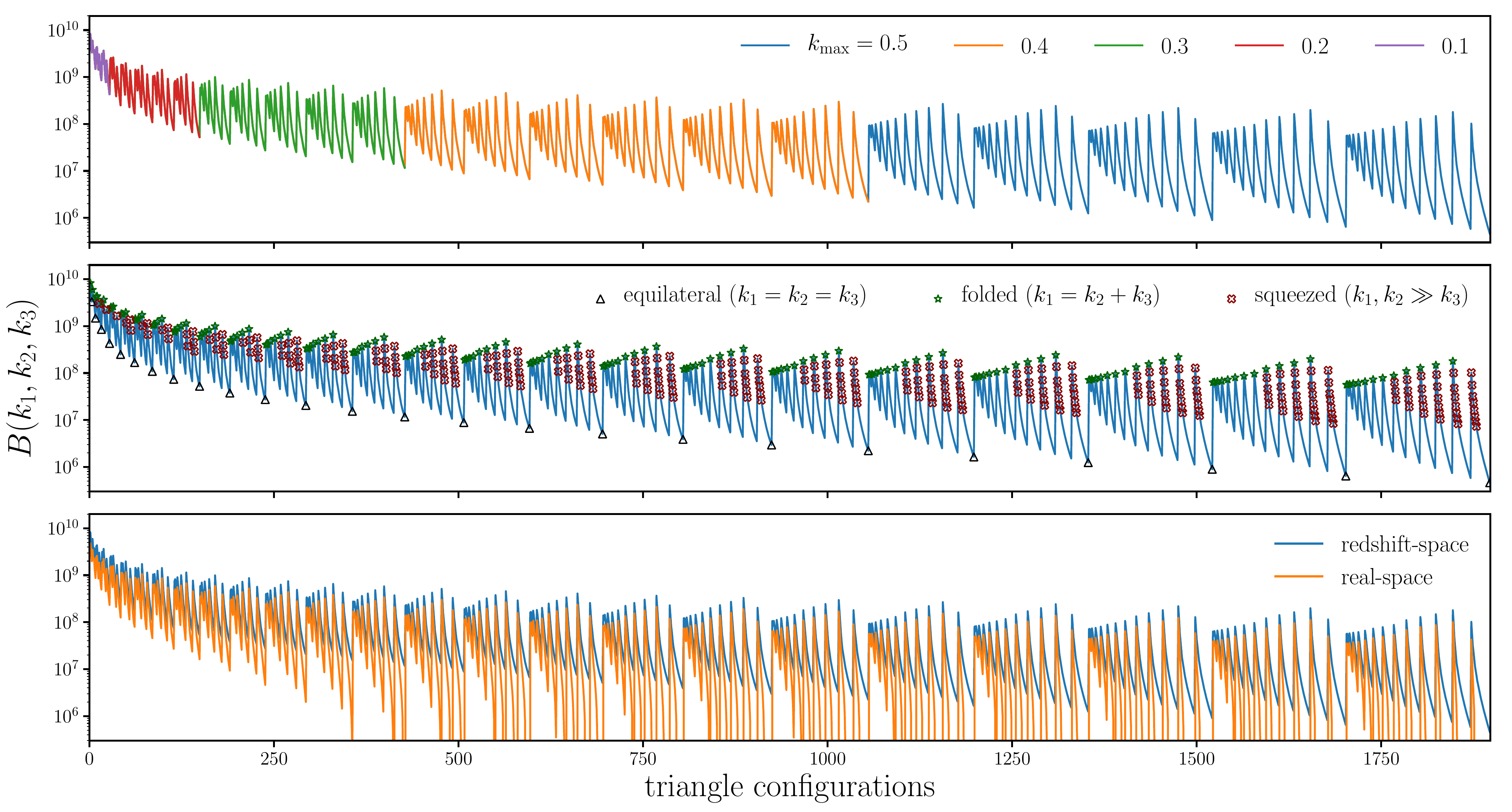}
    \caption{
        {\em Top}: We highlight the triangle configurations of the bispectrum that fall within: 
        $k_{\rm max} <$ 0.1 (purple), 0.2 (red), 0.3 (green), 0.4 (orange), and $0.5~\mpc$ (blue). The number of 
        triangle configurations increase significantly at higher $k_{\rm max}$, which contributes to the 
        significant constraining power on nonlinear scales. 
        {\em Center}: We mark triangle configurations with three different shapes:
        equilateral (black triangle), folded (green star), and squeezed (red cross) triangles. Equilateral 
        triangles have $k_1 = k_2 = k_3$; folded triangles have $k_1 = k_2 + k_3$; squeezed triangles have 
        $k_1, k_2 \gg k_3$, above we mark configurations with $k_1, k_2 > 2.4k_3$ as squeezed. 
        {\em Bottom}: Comparison of the real (orange) and redshift-space (blue) halo bispectrum for 
        fiducial Quijote simulations. The redshift-space $\BOk$ has a significantly higher 
        overall amplitude than the real-space $\BOk$, with a shape and $k_{\rm max}$ dependence. 
        As a result, the redshift-space $\BOk$ has a higher signal-to-noise and more 
        constraining power than the real-space $\BOk$.  
    } 
\label{fig:bk_details}
\end{center}
\end{figure}
\section{Redshift-space Bispectrum} \label{sec:bk_details}
In this work, we present how the redshift-space bispectrum helps break degeneracies 
among cosmological parameters, especially between $\smnu$ and $\sig$, and improve 
constraints on $\smnu$ as well as the other cosmological parameters. We measure 
the redshift-space bispectrum monopole, $\BOk$, of the HADES and Quijote simulations 
(Section~\ref{sec:hades}) using an FFT based estimator similar to the ones in 
\cite{sefusatti2005a}, \cite{scoccimarro2015}, and \cite{sefusatti2016} 
(see Section~\ref{sec:bk} for details). We use triangle configurations defined by 
$k_1$, $k_2$, and $k_3$ bins of width $\Delta k = 3k_f = 0.01885~h/{\rm Mpc}$. 
For $k_{\rm max} = 0.5~h/{\rm Mpc}$ we have $\BOk$ for 1898 
triangle configurations. For $k_{\rm max} = 0.4$, 0.3, 0.2, and $0.1~h/{\rm Mpc}$, 
we have 1056, 428, 150, and 28 triangle configurations respectively. The number
of triangle configurations sharply increases for higher $k_{\rm max}$. This
contributes to the significant increase in constraining power as we go to increasingly 
nonlinear scales (Figure~\ref{fig:fish_kmax}). We highlight the triangle 
configurations at different $k_{\rm max}$ for the $\BOk$ of the Quijote simulation 
at the fiducial cosmology in the top panel of Figure~\ref{fig:bk_details}. We mark 
the configurations within $k_{\rm max} = 0.1, 0.2, 0.3, 0.4$, and $0.5~h/{\rm Mpc}$ 
in purple, red, green, orange, and blue respectively. 

In Figure~\ref{fig:bk_details}, as well as in Figures~\ref{fig:bk_amp}, \ref{fig:dbk_amp}, 
\ref{fig:bk_deriv}, we order the triangle configurations by looping through 
$k_3$ in the inner most loop and $k_1$ in the outer most loop such that 
$k_1 \geq k_2 \geq k_3$. This ordering is different from the ordering in~\cite{gil-marin2017}, 
which loops through $k_3$ in the inner most loop and $k_1$ in the outer most 
increasing loop but with $k_1 \leq k_2 \leq k_3$. As the top panel demonstrates, 
our ordering clearly reflects the $k_{\rm max}$ range of the configurations. 
Furthermore, the repeated configuration sequences in our ordering cycles through
all available triangle configurations for a given $k_1$. In the center panel 
of Figure~\ref{fig:bk_details}, we mark equilateral (black triangle), folded 
(green star), and squeezed (red cross) triangle configurations. Equilateral 
triangles have $k_1 = k_2 = k_3$, folded triangles have $k_1 = k_2 + k_3$, and 
squeezed triangles have $k_1, k_2 \gg k_3$ --- $k_1, k_2 > 2.4k_3$ in Figure~\ref{fig:bk_details}. 
These shapes correspond to the three vertices of the bispectrum shape plots of 
Figures~\ref{fig:bk_shape} and \ref{fig:dbk_shape}. 

Lastly, in the bottom panel of Figure~\ref{fig:bk_details} we compare the 
redshift-space bispectrum (blue) to the real-space bispectrum of the same 
Quijote simulations at the fiducial cosmology (orange). Overall, the 
redshift-space $\BOk$ has a higher amplitude than the real-space $\BOk$ with 
a significant triangle shape dependence: equilateral triangles have the 
largest difference in amplitude while folded triangles have the smallest. 
The relative amplitude difference also depends significantly on $k_{\rm max}$ 
where the difference is larger for configurations with higher $k_1$. With 
its higher amplitude, the redshift-space $\BOk$ has a higher signal-to-noise 
and more constraining power than the real-space $\BOk$.  

%% file: converge.tex
\section{Fisher Forecasts using $N$-body simulations} \label{sec:numerical}
The two key elements in calculating the Fisher matrices in our bispectrum forecasts
are the bispectrum covariance matrix ($\bfi{C}$; Figure~\ref{fig:bk_cov}) and 
the derivatives of the bispectrum along the cosmological and nuisance parameters 
($\partial B_0/\partial \theta$; Figure~\ref{fig:bk_deriv}). We compute 
both these elements directly using the $N$-body simulations of the Quijote suite 
(Section~\ref{sec:hades}). This exploits the accuracy of $N$-body simulations in 
the nonlinear regime and allows us to accurately quantify the constraining power of 
the bispectrum beyond perturbation theory models. However, we must ensure that 
both $\bfi{C}$ and $\partial B_0/\partial \theta$ have converged and that numerical 
effects do not introduce any biases that impact our results. Below, we tests 
the convergence of $\bfi{C}$ and $\partial B_0/\partial \theta$ and discuss
some of the subtleties and caveats of our $\partial B_0/\partial \theta$ 
calculations. 

\begin{figure}
\begin{center}
    \includegraphics[width=0.75\textwidth]{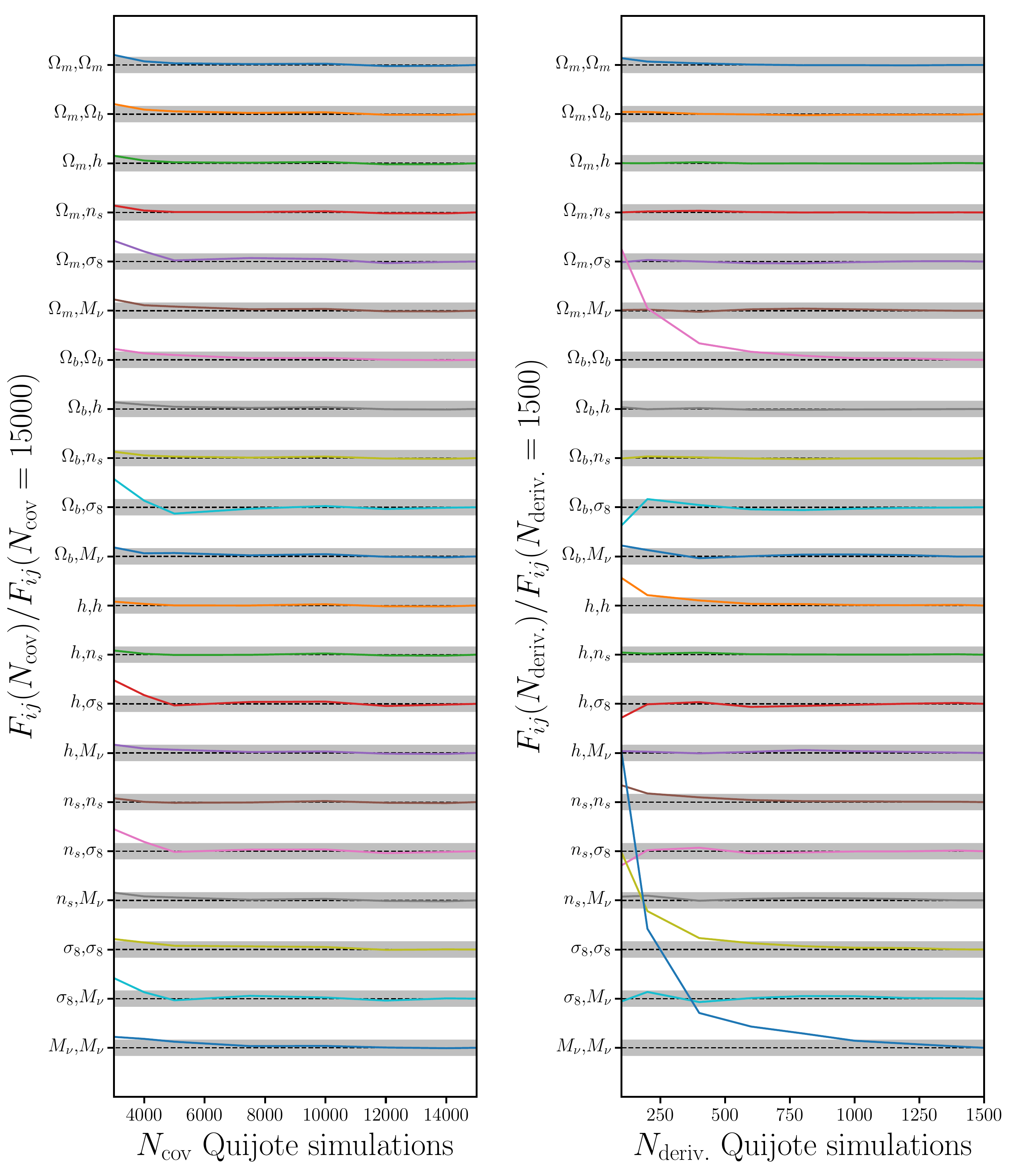}
    \caption{
        {\em Left}: The convergence of all 36 Fisher matrix elements, $F_{ij}$, as 
        as a function of $N_{\rm cov}$, the number of $N$-body simulations used to 
        estimate the covariance matrix, $\bfi{C}$. We present the ratio between 
        $F_{ij}(N_{\rm cov})$, $F_{ij}$ derived from $\bfi{C}$ calculated with 
        $N_{\rm cov}$ simulations, and $F_{ij}(N_{\rm cov} = 15,000)$. We mark 
        $\pm5\%$ deviations in the ratios with the shaded regions for reference. 
        All $F_{ij}$ elements vary by $\lesssim 5\%$ for $N_{\rm cov} > 5000$ and 
        $\lesssim 1\%$ for $N_{\rm cov} > 10,000$.
        {\em Right}: The convergence of $F_{ij}$ as a function of $N_{\rm deriv}$, 
        the number of $N$-body simulations used to estimate the derivatives, 
        $\partial B_0/\partial \theta$. We plot the ratios between $F_{ij}(N_{\rm deriv})$ 
        and $F_{ij}(N_{\rm deriv} = 1,500)$. All $F_{ij}$ elements vary by 
        $\lesssim 5\%$ for $N_{\rm deriv} > 1000$. Hence, {\em $N_{\rm cov} = 15,000$ 
        and $N_{\rm deriv} = 1,500$ are sufficient and the convergence of $\bfi{C}$ 
        and $\partial B_0/\partial \theta$ does not impact $F_{ij}$.}
    }
\label{fig:fij_converge}
\end{center}
\end{figure}

\begin{figure}
\begin{center}
    \includegraphics[width=0.75\textwidth]{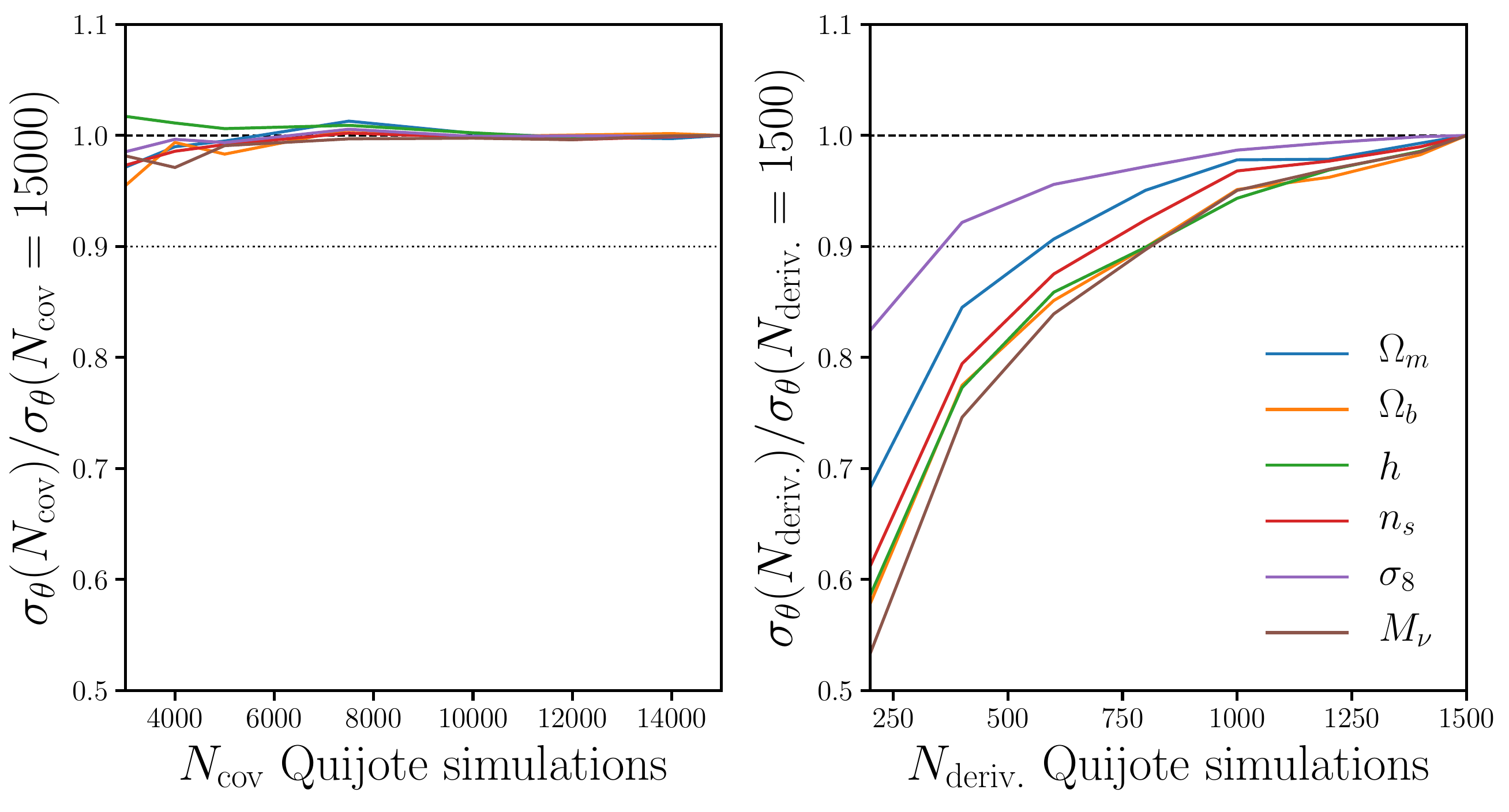} 
    \caption{{\em Left}: The convergence of the marginalized 1$\sigma$ constraints
    for \{$\Om$, $\Ob$, $h$, $n_s$, $\sig$, $\smnu$\}, $\sigma_\theta$, as a function of 
    $N_{\rm cov}$, the number of Quijote simulations used to estimate $\bfi{C}$. We 
    plot the ratio between $\sigma_\theta(N_{\rm cov})$, derived from $\bfi{C}$ with 
    $N_{\rm cov}$ simulations, and $\sigma_\theta(N_{\rm cov} = 15,000)$. $\sigma_\theta$ vary by $\lesssim 5$ and $1\%$ for 
    $N_{\rm cov} > 5000$ and $12,000$, respectively. 
    {\em Right}: The convergence of $\sigma_\theta$ as a function of $N_{\rm deriv}$, 
    the number of simulated used to estimate $\partial B_0/\partial \smnu$. We plot 
    the ratio between $\sigma_\theta(N_{\rm deriv})$, derived from $\partial B/\partial \smnu$
    with $N_{\rm deriv}$ simulations, and $\sigma_\theta(N_{\rm deriv} = 1,500)$. Although 
    $\sigma_\theta(N_{\rm deriv})/\sigma_\theta(N_{\rm deriv} = 1,500)$ vary among 
    the parameters, the ratio vary by $\lesssim 10$ and $5\%$ for $N_{\rm deriv} > 800$ 
    and $1200$, respectively. Hence, {\em we have a sufficient number of simulations 
    to estimate $\bfi{C}$ and the derivatives of the bispectrum and our forecasts are 
    robust to their convergence.} 
    }
\label{fig:converge}
\end{center}
\end{figure}

We use 15,000 Quijote $N$-body simulations at the fiducial cosmology to 
estimate $\bfi{C}$. This is a \emph{significantly} larger number of simulations 
than any previous bispectrum analyses; however, we also consider 1898 triangle 
configurations out to $k_{\rm max} = 0.5~\mpc$. For reference, \cite{gil-marin2017} 
recently used 2048 simulations to estimate the covariance matrix of the 
bispectrum with $825$ configurations. To check the convergence of covariance 
matrix, we vary the number of simulations used to estimate $\bfi{C}$, 
$N_{\rm cov}$, and determine whether this significantly impacts the elements 
of the Fisher matrix, $F_{ij}$, or the final marginalized Fisher parameter 
constraints, $\sigma_\theta$. In the left panel of Figure~\ref{fig:fij_converge}, 
we present the ratio between $F_{ij}(N_{\rm cov})$, $F_{ij}$ derived from 
$\bfi{C}$ calculated with $N_{\rm cov}$ simulations, and 
$F_{ij}(N_{\rm cov} = 15,000)$ for all 36 elements of the Fisher matrix. We 
shade $\pm5\%$ deviations in the ratios for reference. The $F_{ij}$ elements 
vary by $\lesssim 5\%$ for $N_{\rm cov} > 5000$ and $\lesssim 1\%$ for 
$N_{\rm cov} > 10,000$. Next, we present ratio between $\sigma_\theta(N_{\rm cov})$, 
the marginalized $1\sigma$ constraints for $\{\Om$, $\Ob$, $h$, $n_s$, $\sig$, $\smnu\}$ 
derived from $\bfi{C}$ calculated with $N_{\rm cov}$ simulations, and 
$\sigma_\theta(N_{\rm cov} = 15,000)$ in the left panel of Figure~\ref{fig:converge}. 
The constraints vary by $\lesssim 5\%$ for $N_{\rm cov} > 5000$ and $\lesssim 1\%$ 
for $N_{\rm cov} > 12000$. Hence, $N_{\rm cov} = 15,000$ is sufficient to 
accurately estimate $\bfi{C}$ and its convergence does not impact our 
forecasts.

We estimate $\partial B_0/\partial \theta$ using $N_{\rm deriv} = 1,500$ 
$N$-body simulations at 13 different cosmologies listed in Table~\ref{tab:sims} 
To check the convergence of $\partial B_0/\partial \theta$ and its impact on our 
results, we examine the ratio between $F_{ij}(N_{\rm deriv})$, the Fisher 
matrix element derived from $\partial B_0/\partial \theta$ calculated with 
$N_{\rm deriv}$ simulations, and $F_{ij}(N_{\rm deriv} = 1,500)$ for all 
36 elements of the Fisher matrix in the right panel of Figure~\ref{fig:fij_converge}.
For $N_{\rm deriv} > 1000$, $F_{ij}$ elements vary by $\lesssim 5\%$. Next, we
present the ratio between $\sigma_\theta(N_{\rm deriv})$, the marginalized 
$1\sigma$ constraints for $\{\Om$, $\Ob$, $h$, $n_s$, $\sig$, $\smnu\}$ 
derived from $\partial B_0/\partial \theta$ calculated with $N_{\rm deriv}$ 
simulations, and $\sigma_\theta(N_{\rm deriv} = 1,500)$ in the right panel 
of Figure~\ref{fig:converge}. Unlike $\sigma_\theta(N_{\rm cov})$, 
$\sigma_\theta(N_{\rm deriv})$ depend significantly on $\theta$. For instance, 
$\sigma_{\sig}$ and $\sigma_{\Om}$ vary by $\lesssim 10\%$ for $N_{\rm deriv} > 600$ 
and $\lesssim 1\%$ for $N_{\rm deriv} > 1200$. $\sigma_\theta$ for the other 
parameter vary significantly more. Nonetheless, for $N_{\rm deriv} > 800$ and 
$1200$ they vary by $\lesssim 10$ and $5\%$, respectively.

\begin{figure}
\begin{center}
    \includegraphics[width=\textwidth]{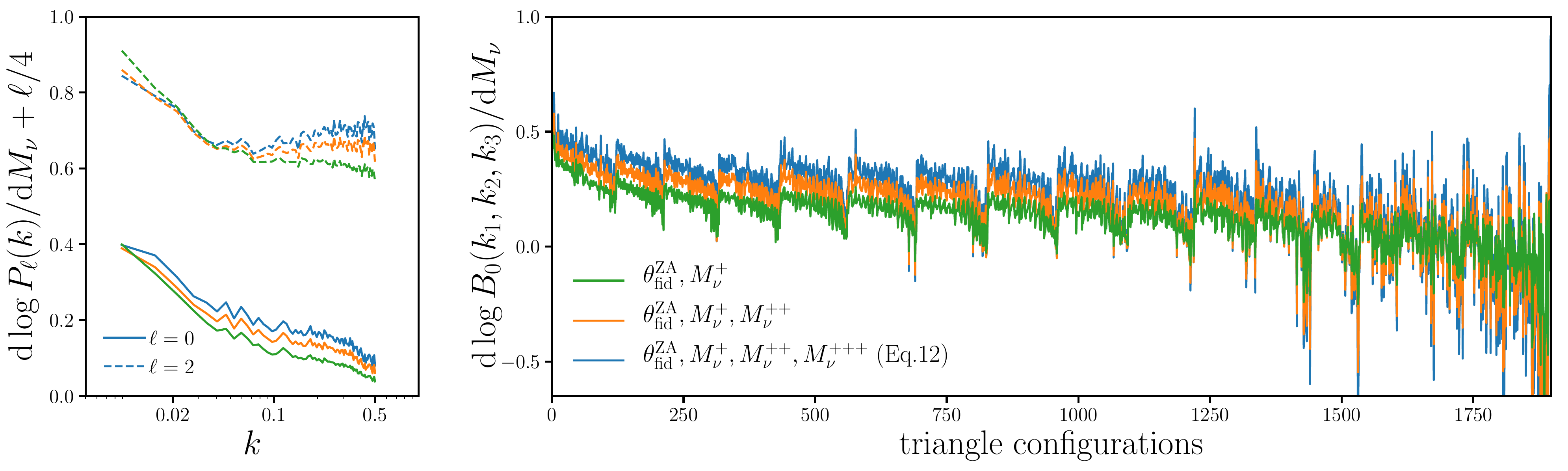} 
    \caption{Comparison of $\partial \log B_0(k_1, k_2, k_3)/\partial M_\nu$ (right) 
    and $\partial \log P_\ell(k)/\partial M_\nu$ (left), computed using Quijote simulations 
    at \{$\theta_{\rm fid}^{\rm ZA}$, $\smnu^{+}$, $\smnu^{++}$, $\smnu^{+++}$\} (blue), 
    \{$\theta_{\rm fid}^{\rm ZA}$, $\smnu^{+}$, $\smnu^{++}$\} (orange), and \{$\theta_{\rm fid}^{\rm ZA}$, $\smnu^{+}$\} 
    (green). The derivative approximations differ from one another by $\sim10\%$ with 
    Eq.~\ref{eq:dbkdmnu} producing the largest estimate for both $P_\ell$ and $B_0$.
    Using the \{$\theta_{\rm fid}^{\rm ZA}$, $\smnu^{+}$, $\smnu^{++}$\} derivatives instead of the 
    Eq.~\ref{eq:dbkdmnu} derivatives, increases the marginalized constraint on $\smnu$ 
    by $\sim 30\%$. However, the differences in the derivatives propagate similarly to 
    the $P_\ell$ and $B_0$ forecasts so the relative improvement of $B_0$ over $P_\ell$
    remains the same. Hence, we conclude that {\em the derivatives with respect to $\smnu$ 
    are sufficiently stable and robust for our Fisher forecasts}.
    }
\label{fig:dPBdmnu}
\end{center}
\end{figure}

For $\Om$, $\Ob$, $h$, $n_s$, $\sig$, and also the nuisance parameter $M_{\rm lim}$ 
we estimate $\partial B_0/\partial \theta$ using a centered difference approximation 
(Eq.~\ref{eq:dbkdt}). However, for $\smnu$ we cannot have values below 0.0 eV 
and, thus, cannot estimate the derivative with the same method. If we use the 
forward difference approximation, 
\beq 
\frac{\partial \overline{B}_0}{\partial \smnu} \approx \frac{\overline{B}_0(\theta_{\rm fid}^{\rm ZA}+\delta \smnu) - \overline{B}_0(\smnu^{\rm fid})}{\delta \smnu}, 
\eeq
the error goes as $\mathcal{O}(\delta \smnu)$. Instead, we use a finite difference
approximation with the Quijote simulations at $\smnu^{+}$, $\smnu^{++}$, $\smnu^{+++}$, 
and the fiducial cosmology, a $\mathcal{O}(\delta \smnu^2)$ order approximation 
(Eq.~\ref{eq:dbkdmnu}). We can also use a finite difference approximation with 
simulations at $\smnu^{+}$, $\smnu^{++}$, and the fiducial cosmology. We examine
the stability of $\partial \log B_0/\partial M_\nu$ (right) and 
$\partial \log P_\ell/\partial M_\nu$ (left) by comparing the derivatives computing
using simulations at \{$\theta_{\rm fid}^{\rm ZA}$, $\smnu^{+}$, $\smnu^{++}$, $\smnu^{+++}$\} (blue), 
\{$\theta_{\rm fid}^{\rm ZA}$, $\smnu^{+}$, $\smnu^{++}$\} (orange), and \{$\theta_{\rm fid}^{\rm ZA}$, $\smnu^{+}$\} 
(green) in Figure~\ref{fig:dPBdmnu}. The three $\partial \log B_0/\partial \smnu$
approximations differ from one another by $\sim10\%$ with Eq.~\ref{eq:dbkdmnu} 
producing the largest estimate for both $P_\ell$ and $B_0$. If we use the
\{$\theta_{\rm fid}^{\rm ZA}$, $\smnu^{+}$, $\smnu^{++}$\} derivative and 
\{$\theta_{\rm fid}^{\rm ZA}$, $\smnu^{+}$\} derivative 
instead of Eq.~\ref{eq:dbkdmnu} for our Fisher forecasts, the marginalized constraint on
$\smnu$ for $k_{\rm max} = 0.5~\mpc$ increases to 0.390 and 0.682 eV for $P_\ell$ 
and 0.0754 and 0.1354 eV for $B_0$. Compared to our $\sigma_{\smnu} = 0.0572$ eV 
$B_0$ forecast, these correspond to a $\sim30$ and $130\%$ relative increase. 
While the derivative estimated from \{$\theta_{\rm fid}^{\rm ZA}$, $\smnu^{+}$\} significantly 
impact the forecasts, we emphasize that this is a $\mathcal{O}(\delta \smnu)$ approximation, 
unlike the other $\mathcal{O}(\delta \smnu^2)$ approximations. In fact, the 
\{$\theta_{\rm fid}^{\rm ZA}$, $\smnu^{+}$\} derivatives are better $\mathcal{O}(\delta \smnu^2)$ 
estimates for $\partial P_\ell/\partial \smnu$ and $\partial B_0/\partial \smnu$ at 0.05 eV. 
When we compare the derivative of the linear theory power spectrum, $P^{(\rm LT)}$, 
we find that $\partial P^{(\rm LT)}/\partial \smnu$ at 0.0 is larger than at 0.05 eV. 
Hence, the differences between the \{$\theta_{\rm fid}^{\rm ZA}$, $\smnu^{+}$\} derivatives 
and the other derivatives are not solely due to numerical stability. Moreover, because the 
discrepancies in the derivative propagate similarly to both $P_\ell$ and $B_0$ forecasts, 
the relative improvement of $B_0$ over $P_\ell$ remains roughly the same.
Hence, we conclude that the derivatives with respect to $\smnu$ are sufficiently 
stable and robust for our Fisher forecasts.

